%% file: main_Clean.tex
\documentclass[journal,twoside,web]{ieeecolor2}
\usepackage{sty/generic}
\usepackage{cite}
\usepackage{amsmath,amssymb,amsfonts}
\usepackage{algorithmic}
\usepackage{graphicx}
\usepackage{textcomp}

\makeatletter
\usepackage[table]{xcolor}
\usepackage{xspace}
\def\@onedot{\ifx\@let@token.\else.\null\fi\xspace}
\DeclareRobustCommand\onedot{\futurelet\@let@token\@onedot}

\def\ie{\emph{i.e}\onedot}

\usepackage{adjustbox}
\usepackage{xargs}  
\usepackage[dvipsnames]{xcolor}
\usepackage[colorinlistoftodos,prependcaption,textsize=tiny]{todonotes}
\definecolor{revred}{RGB}{190,80,90}
\newcommandx{\firstRev}[2][1=]{\todo[linecolor=OliveGreen,backgroundcolor=OliveGreen!25,bordercolor=OliveGreen,#1]{#2}}
\newcommandx{\secondRev}[2][1=]{\todo[linecolor=cyan,backgroundcolor=cyan!25,bordercolor=cyan,#1]{#2}}
\newcommandx{\thirdRev}[2][1=]{\todo[linecolor=Blue,backgroundcolor=Blue!25,bordercolor=Blue,#1]{#2}}
\newcommandx{\forthRev}[2][1=]{\todo[linecolor=revred,backgroundcolor=revred!20,bordercolor=revred,#1]{#2}}
\newcommandx{\revAE}[2][1=]{\todo[linecolor=BurntOrange,backgroundcolor=BurntOrange!25,bordercolor=BurntOrange,#1]{#2}}

\usepackage{multirow}
\usepackage{verbatim}
\definecolor{utku_green}{RGB}{0,153,0}
\usepackage[]{color-edits}
\addauthor{ma}{red}
\addauthor{ms}{blue}

\definecolor{mildyellow}{RGB}{255, 245, 180}
\definecolor{custom_velvet}{RGB}{148,0,64}
\definecolor{custom_green}{RGB}{0,148,64}
\definecolor{subsectionblue}{rgb}{0,0.263,0.576}

\usepackage{booktabs}
\usepackage{makecell}
\usepackage{multirow}
\usepackage{amssymb}
\usepackage{lipsum}
\usepackage{hyperref}
\usepackage{cleveref}
\crefname{figure}{Fig.}{Figs.}
\Crefname{equation}{Eq.}{Eqs.}
\Crefname{table}{Tab.}{Tabs.}
\crefname{figure}{Fig.}{Figs.}
\Crefname{equation}{Eq.}{Eqs.}
\Crefname{table}{Tab.}{Tabs.}
\crefname{section}{Sec.}{Secs.}

\def\BibTeX{{\rm B\kern-.05em{\sc i\kern-.025em b}\kern-.08em
    T\kern-.1667em\lower.7ex\hbox{E}\kern-.125emX}}
\begin{document}

\title{Harnessing Magnitude-Only and Complex Measurements for Improved Dynamic MRI Reconstruction with Learned Priors} 
\author{Mahdi Saberi, \IEEEmembership{Student Member, IEEE}, Ya\c sar Utku Al\c  calar, \IEEEmembership{Student  Member, IEEE}, Merve G\"{u}lle, \IEEEmembership{Student Member, IEEE}, Chetan Shenoy, and Mehmet Ak\c cakaya, \IEEEmembership{Senior Member, IEEE}
\thanks{Manuscript submitted June 1, 2026. This work was partially supported by NIH R01HL153146, NIH R01HL179616, NIH R01EB032830, NIH P41EB027061. \emph{(Corresponding author: Mehmet~Ak\c{c}akaya)}}
\thanks{Mahdi Saberi, Ya\c{s}ar~Utku~Al\c{c}alar, Merve G\"{u}lle and Mehmet Ak\c{c}akaya are with Department of Electrical and Computer Engineering, University of Minnesota, Minneapolis, MN, USA, and with Center for Magnetic Resonance Research, University of Minnesota, Minneapolis, MN, USA (e-mail: {saber032@umn.edu, alcal029@umn.edu, glle0001@umn.edu, akcakaya}@umn.edu). Chetan Shenoy is with Department of Medicine (Cardiology), University of Minnesota, Minneapolis, MN, USA (e-mail: cshenoy@umn.edu).}
}

\maketitle

\begin{abstract}
MRI reconstruction methods for undersampled k-space data naturally utilize complex-valued measurements. Parallel developments in sparse phase retrieval have shown that magnitude-only measurements may provide complementary information for signal recovery. However, their use in MRI reconstruction remains largely unexplored, due to lack of practical settings where informative magnitude measurements can be obtained without additional scan time. 
In this work, we investigate the use of auxiliary k-space magnitude information for accelerated steady-state dynamic MRI reconstruction, and demonstrate strong consistency of k-space magnitudes across time-frames. Building on this observation, we propose $\mathbb{C}+\text{Mag}$, a magnitude-informed physics-driven deep learning (PD-DL) reconstruction method. The proposed method employs an ADMM-based unrolling framework with a novel magnitude-aware data-fidelity formulation, where quadratically smoothed optimization and momentum-based updates are introduced to address the non-differentiability and non-convexity of the magnitude constraints. Experiments on retrospectively undersampled cine MRI and phase-contrast flow MRI datasets, as well as prospectively undersampled real-time cine MRI acquisitions, demonstrate improved artifact suppression, sharper anatomical recovery, and better preservation of phase information compared to conventional PD-DL methods, which is further supported through blinded expert reader evaluations. 
\end{abstract}


\bstctlcite{IEEEexample:BSTcontrol}

\section{Introduction}
\label{sec:introduction}
\IEEEPARstart{R}{educing} MRI acquisition time often requires recovering images from highly undersampled k-space data, resulting in a severely ill-posed inverse problem. Classical compressed sensing (CS) approaches address this challenge through sparsity-promoting regularization~\cite{lustig2007sparse}, while recent physics-driven deep learning (PD-DL) methods combine acquisition physics with learned priors to achieve substantially improved reconstruction quality~\cite{yang2020admm_csnet,knoll2020deep-survey,hammernik2023SPM}. This area has seen a number of developments, including more expressive regularizers \cite{kabas2026physics, junno2025_TE-MRI_NIPS,gao2021accelerating,saberi2026umpire, Korkmaz2022}, newer learning strategies \cite{akcakaya2022_SPMsurvey, alcalar2025_SPIC_SSDU}, as well as diffusion model based solvers \cite{jalal2021robust, chung2022scoreMRI, askari2025training, jiang2024fast, gulle2026_PnP-CM_CVPR}.
 Nevertheless, reconstruction performance at high acceleration rates remains fundamentally constrained by the limited amount of acquired information, frequently leading to residual artifacts, anatomical blurring, or loss of fine structural detail.

In parallel, the computational imaging community has extensively studied phase retrieval problems, where the goal is to recover signals from magnitude-only measurements~\cite{wang2017sparta,chung2023dps,gulle2026_PnP-CM_CVPR}. In particular, theoretical studies have shown that, under suitable conditions, two magnitude-only measurements can carry information equivalent to one complex linear measurement in sparse recovery problems~\cite{akccakaya2015sparse}, motivating the use of combined complex-valued and magnitude-only measurements for MRI reconstruction~\cite{akcakaya2015joint}. However, despite this theoretical promise, the use of auxiliary magnitude information in MRI reconstruction has remained largely underexplored~\cite{park2017sms} due to two reasons: i) lack of practical acquisition settings where informative magnitude-only measurements can be obtained without increasing scan time, ii) limited improvement from using hand-designed regularizers in these settings.

The recent advent of large-scale raw MRI databases has renewed interest in understanding the structure of k-space magnitude across distinct datasets, albeit not for reconstruction. Studies in Fourier-domain representation learning~\cite{yang2020fda} suggest that the magnitude and phase components capture low-level distributional features and high-level semantic information respectively. This idea was used in the backdoor attack literature~\cite{feng2022fiba} to devise triggers for magnitude k-space, leading to visually imperceptible attacks. Building on these observations, we revisit the utility of using additional magnitude-only k-space measurements in MRI reconstruction. In particular, using large-scale raw MRI databases, we investigate the consistency of multi-coil k-space magnitude across cardiac phases,demonstrating consistently high cosine similarity between neighboring temporal frames for various steady-state dynamic MRI applications. 

Motivated by these findings, we propose a physics-driven deep learning framework for MRI reconstruction using a combination of complex-valued and magnitude-only measurements, referred to as {$\mathbb{C}+\text{Mag}$} reconstruction. Specifically, we formulate a new objective function for MRI reconstruction with joint complex and magnitude least-squares data fidelity terms, where the latter uses auxiliary magnitude constraints derived from neighboring cardiac phases.
The resulting objective is solved using algorithm unrolling with learned proximal operators and a \emph{novel} data-fidelity formulation tailored for mixed complex and magnitude measurements. As the magnitude fidelity term is non-differentiable and non-convex, we further develop a quadratically smoothed formulation and investigate momentum-based optimization strategies within the data-fidelity updates. Our key contributions include:
\begin{itemize}
\item We introduce a PD-DL reconstruction framework that jointly leverages complex-valued and auxiliary magnitude-only measurements for accelerated MRI reconstruction through a unified optimization formulation.
\item We develop an ADMM-based unrolled reconstruction algorithm with a novel magnitude-aware data-fidelity formulation, including quadratically smoothed optimization and momentum-based updates for improved stability. 
\item We conduct extensive evaluations on retrospectively undersampled cine MRI~\cite{chen2020ocmr} and phase-contrast flow MRI datasets~\cite{wang2024cmrxrecon}, as well as a local prospectively undersampled real-time cine MRI databases, demonstrating substantial improvements over conventional PD-DL reconstruction methods at high acceleration rates.
\item We further validate the proposed framework through blinded cardiologist assessments and comprehensive ablation studies analyzing the major design choices of the proposed method.
\end{itemize}

This work is an invited extended submission of our IEEE ISBI paper~\cite{saberi2026_ISBI_CMag}, which introduced the $\mathbb{C}+\text{Mag}$ formulation and established its feasibility for retrospectively undersampled cine MRI at a single acceleration setting. The present manuscript substantially expands upon that preliminary study in several directions. First, we introduce additional optimization formulations and systematically analyze their computational efficiency and convergence behavior. Second, we evaluate the compatibility of the proposed magnitude-informed data-fidelity formulation with both spatial and spatiotemporal learned regularizers. Third, we perform extensive ablation studies on key algorithmic and design choices, including the smoothing strategy for the magnitude term, the number of unrolled and inner data-fidelity iterations, momentum-based data-fidelity solvers, and the amount, temporal distribution, and spatial-frequency content of the auxiliary magnitude information. The experimental scope is also considerably broadened to include multiple acceleration factors, phase-contrast Flow2D MRI, multi-field-strength cine data, and prospective real-time cine acquisitions. Finally, the clinical relevance of the proposed framework is assessed through blinded cardiologist evaluations, cardiac volumetric analysis, and quantitative flow measurements, providing a more practical assessment of whether the observed reconstruction improvements translate into clinically meaningful image quality and functional measurements. Collectively, these extensions transform the original proof-of-concept into a substantially more comprehensive assessment of the generality, computational characteristics, robustness, and clinical utility of the proposed framework.

\section{Background \& Motivation}
\subsection{Background on PD-DL Reconstruction in MRI}\label{sec:background}
Canonical regularized MRI reconstruction solves:
\begin{equation}\label{eq:inverse_typical}
    \arg \min_{\mathbf{x}} \|\mathbf{y}_\Omega -\mathbf{E}_\Omega \mathbf{x} \|_2^2 +\mathcal{R}(\mathbf{x}) 
\end{equation}
\looseness = -1
where $\mathbf{y}_\Omega \in {\mathbb C}^m$ is the undersampled multi-coil k-space measurements acquired with sampling pattern $\Omega$, $\mathbf{E}_\Omega: {\mathbb C}^n \to {\mathbb C}^m$ is the corresponding multi-coil encoding operator, and $\mathcal{R}(\cdot)$ is a regularizer. The first quadratic term enforces data fidelity (DF) with the acquired complex-valued measurements, while the regularizer incorporates prior information. PD-DL methods unroll an iterative algorithm for solving this regularized least squares problem~\cite{fessler2020SPM} for a fixed number of steps, where the DF unit is implemented by conventional methods with learnable parameters, and the proximal operator for the regularizer is implemented implicitly using a neural network~\cite{aggarwal2019MoDL, hammernik2023SPM, hammernik2018VarNet}. The unrolled network can then be trained using either supervised~\cite{aggarwal2019MoDL, hammernik2018VarNet} or self-supervised learning~\cite{yaman2020SSDU, akcakaya2022_SPMsurvey, saberi2026phase}. 

In this work, our primary focus is not the training paradigm itself, but rather the underlying inverse problem formulation and the corresponding optimization strategy. Nevertheless, both supervised and self-supervised training settings are considered in the retrospective and prospective undersampling experiments, respectively.

\begin{figure}[!t]
  \centering
  \includegraphics[width=\linewidth]{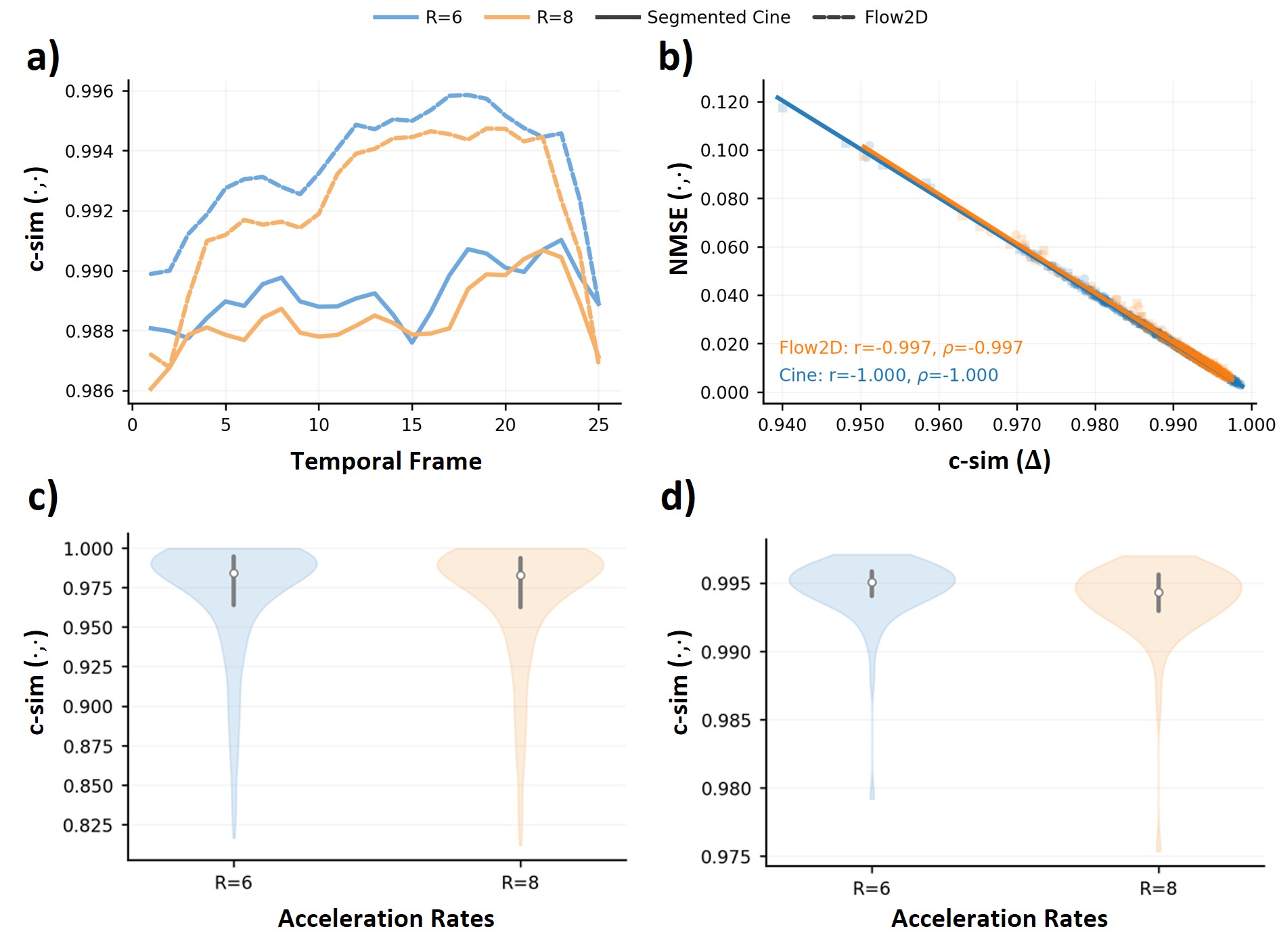}
  \caption{Temporal consistency and estimation error analysis of the auxiliary k-space magnitude information on $\Delta = \Omega^C$.
(a) Mean temporal c-sim across all subjects and slices for segmented cine and Flow2D at $\mathrm{R}\in\{6,8\}$. 
(b) Relationship between NMSE and c-sim, howing a strong inverse correlation for both datasets.
(c,d) Coil-wise c-sim distributions for segmented cine and Flow2D, respectively, at $\mathrm{R}\in\{6,8\}$, where the white markers indicate the median and the bars denote the interquartile range.}
  \label{fig:Motivation}
  \vspace{-2ex}
\end{figure}

\begin{figure*}[!t]
  \centering
  \includegraphics[width=\linewidth]{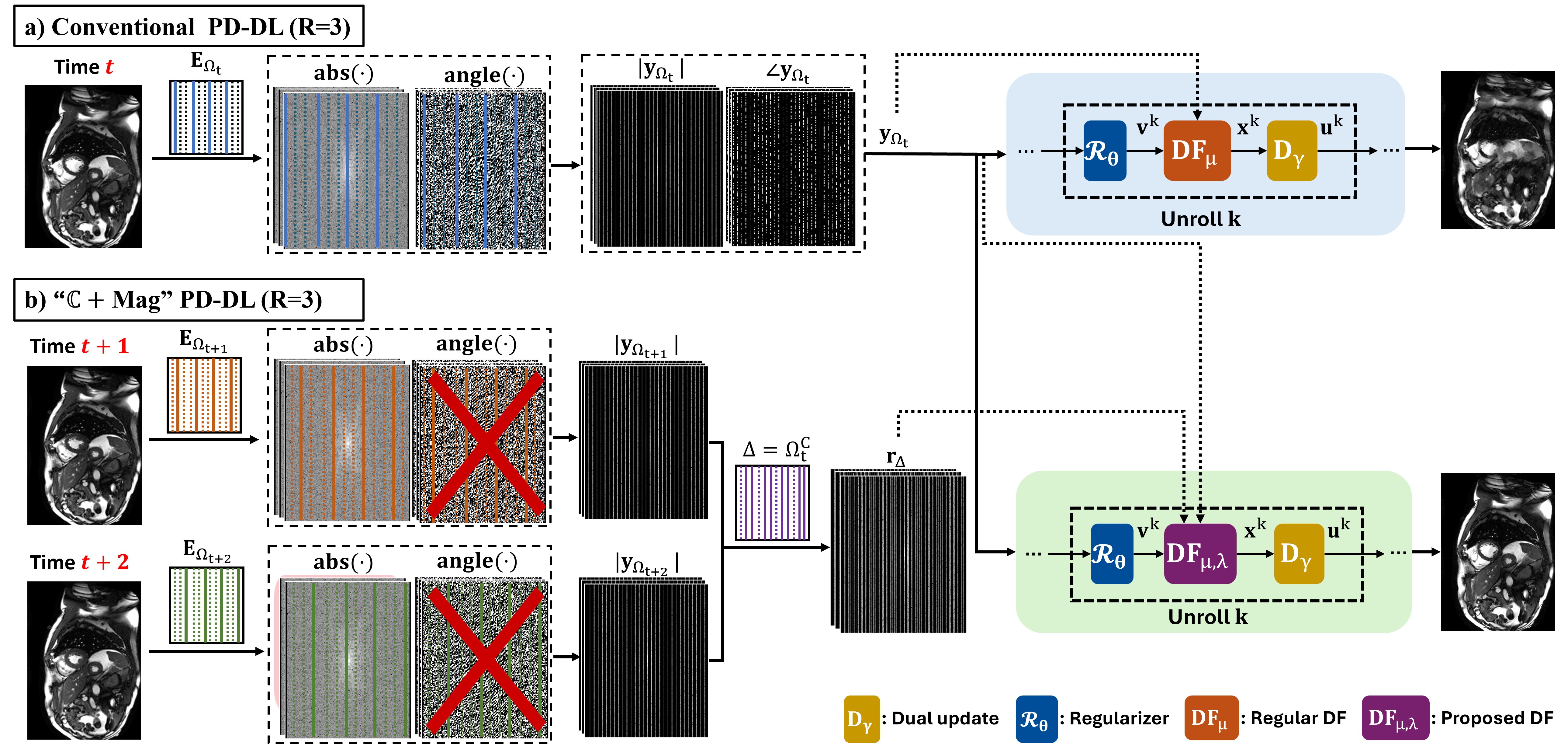}
  \caption{ $\mathbb{C} + \mathrm{Mag}$ reconstruction pipeline that integrates both complex-valued and magnitude-only measurements. Acceleration rate {\unboldmath $\mathrm{R} = 3$} is used for illustration, without loss of generality. For the cardiac phase of interest {\unboldmath $t$}, the complex-valued measurements, {\unboldmath ${\bf y}_{\Omega_t}$}, are used for reconstruction, both in regular PD-DL and in proposed $\mathbb{C} + \mathrm{Mag}$. From adjacent cardiac phases {\unboldmath $t+1$} and {\unboldmath $t+2$}, only the magnitude of the k-space data on lines specified by {\unboldmath $\Omega_{t+1}$} and {\unboldmath $\Omega_{t+2}$} are used as auxiliary information in $\mathbb{C} + \mathrm{Mag}$, denoted by {\unboldmath ${\bf r}_{\Delta}$} with {\unboldmath $\Delta = \Omega_{t+1} \cup \Omega_{t+2} = \Omega_t^C,$} and {\unboldmath $C$} denotes the complement set. Note {\unboldmath ${\bf r}_{\Delta}$} is readily available without any changes to the cine MRI acquisition. Learnable parameters corresponding to each unroll are highlighted using subscripts. }
  \vspace{-3ex}
  \label{fig:pipeline}
\end{figure*}

\subsection{k-space Magnitude Similarity}
Joint sparse phase retrieval and CS theory suggests that two random magnitude measurements holds the same information as one single random complex-valued linear measurement~\cite{akccakaya2015sparse}. Therefore, in scenarios where additional magnitude-only measurements are available (\ie, \emph{accurate} magnitude information with incorrect phase), one can leverage this information to improve MRI reconstruction quality. Prior work has explored several simple scenarios of this setting on a limited number of examples with hand-crafted regularizers~\cite{akcakaya2015joint}.

Motivated by the availability of large-scale MRI databases and recent findings in adversarial robustness literature~\cite{feng2022fiba}, we investigate whether auxiliary magnitude information can be reliably leveraged for MRI reconstruction without requiring additional scan time. In particular, steady-state dynamic MRI provides a natural setting for this formulation due to the high consistency of their multi-coil k-space magnitudes across time.

To quantify this consistency, we focus on breath-hold segmented cine acquisitions from the OCMR dataset~\cite{chen2020ocmr} and phase-contrast Flow2D acquisitions from the CMRxRecon2025 challenge dataset~\cite{cmrxrecon2025}, and analyze the cosine similarity of multi-coil k-space magnitudes across cardiac phases over multiple subjects and slices in the test sets, evaluated over the complementary locations $\Omega^{C}$:
\begin{equation}
    \mathrm{c\text{-}sim}(\lvert\mathbf{y}\rvert,\mathbf{r})
    =
    \frac{
        \left\langle \lvert\mathbf{y}\rvert,\mathbf{r} \right\rangle
    }{
        \left\|\lvert\mathbf{y}\rvert\right\|
        \left\|\mathbf{r}\right\|}.
\end{equation}
where $\mathbf{y}$ denotes the target k-space, $\mathbf{r}$ denotes the magnitude estimate obtained from adjacent cardiac phases, and $|\cdot|$ denotes the element-wise absolute value operator. 
\Cref{fig:Motivation}(a) shows the c-sim remains consistently high across cardiac phases over $\Omega^C$ for both datasets at $\mathrm{R}\in\{6,8\}$. While c-sim characterizes the overall alignment of magnitude vectors, it does not directly quantify their relative error. We therefore additionally consider the normalized magnitude estimation error (NMSE). In particular, the two quantities are related, where defining $\mathrm{q}\triangleq  \|\mathbf{r}\|_2 / \||\mathbf{y}|\|_2$, yields: 
\begin{align}
    \mathrm{NMSE}(\lvert\mathbf{y}\rvert,\mathbf{r}) =  \frac{\|\mathbf{r}-|\mathbf{y}|\|_2^2}{\||\mathbf{y}|\|_2^2} =
    1+\mathrm{q}^2-2\mathrm{q}\cdot\mathrm{c\text{-}sim}.
\end{align}
Thus, for a comparable energy (\ie $q\approx1$), 
 $   \text{NMSE}
    \approx
    2\left(1-\mathrm{c\text{-}sim}\right)
$.
\Cref{fig:Motivation}(b) shows this strong inverse relationship between c-sim and NMSE, where high c-sim corresponds to low NMSE at the unsampled k-space locations for both datasets. To further characterize the reliability of the auxiliary magnitude information, we examined its coil-wise behavior for Flow2D MRI dataset. As shown in \Cref{fig:Motivation}(c,d), c-sim remains consistently high across coils with limited variability, in agreement with the fully vectorized multi-coil analysis.
\input{Tables/acqusition_params}
\section{Methods}\label{sec:Method}
These findings motivate a reconstruction framework that jointly enforces consistency with both the acquired complex-valued measurements and auxiliary magnitude constraints, such as those from neighboring timeframes. Accordingly, we formulate accelerated MRI reconstruction using both complex-valued and magnitude-only measurements as the following regularized optimization problem:
\begin{equation}\label{eq:inverse}
    \begin{aligned}
        \arg \min_{\mathbf{x}} \|\mathbf{y}_\Omega - \mathbf{E}_\Omega \mathbf{x} \|_2^2  + \lambda\|\mathbf{r}_\Delta - |\mathbf{E}_\Delta \mathbf{x}| \|_2^2 + \mathcal{R}(\mathbf{x})
    \end{aligned}
\end{equation}
where $\lambda$ is a weight term, ${\bf y}_\Omega$ is the complex-valued measurements as before, and ${\bf r}_\Delta$ is the auxiliary magnitude-only measurements available on locations indexed by $\Delta$. In this context, $\Delta \subseteq \Omega^C$, ideally with $\Delta = \Omega^C,$ such that the auxiliary magnitude data provides information on all k-space locations that are not directly sampled in the target time-frame. In many applications, this can be naturally realized using shifted interleaved sampling trajectories across neighboring temporal frames, as commonly employed in real-time imaging~\cite{kellman2001TSENSE}. The effects of the number and temporal distribution of auxiliary neighboring frames, as well as their temporal distance from the target frame, are systematically investigated in~\Cref{sec:AblationMagEst}.

We solve the objective in \eqref{eq:inverse} using ADMM \cite{sun2016deep, gu2022revisiting, saberi2026umpire}, leading to the following subproblems: 
\begin{align}
\mathbf{v}^{\mathrm{k}} 
    &= \arg \min_{\mathbf{v}} \;
    \mu \|\mathbf{x}^{\mathrm{k}-1} - \mathbf{v} + \mathbf{u}^{\mathrm{k}-1}\|_2^2 
    + \mathcal{R}(\mathbf{v})
    \label{eq:v_update} \\ 
\mathbf{x}^{\mathrm{k}} 
    &= \arg \min_{\mathbf{x}} \;
    \|\mathbf{y}_\Omega - \mathbf{E}_\Omega \mathbf{x}\|_2^2
    + \lambda \|\mathbf{r}_\Delta - \lvert \mathbf{E}_\Delta \mathbf{x} \rvert\|_2^2
    \nonumber \\
    &+ \mu \|\mathbf{x} - \mathbf{v}^{\mathrm{k}} + \mathbf{u}^{\mathrm{k}-1}\|_2^2
    \label{eq:x_update} \\       
\mathbf{u}^{\mathrm{k}} 
    &= \mathbf{u}^{\mathrm{k}-1} 
    + \gamma(\mathbf{x}^{\mathrm{k}} - \mathbf{v}^{\mathrm{k}})
    \label{eq:u_update}.
\end{align}
where $\mu, \lambda, \gamma$ are learnable parameters. Similar to conventional PD-DL~\cite{hammernik2023SPM}, \Cref{eq:v_update} is implicitly implemented using a neural network, while ~\Cref{eq:x_update} is solved iteratively itself. To this end, the derivative of the objective function in \eqref{eq:x_update} needs to be taken with respect to $\bar{\mathbf{x}}$ based on CR-Calculus~\cite{kreutz2009complex}. However, the DF step in ~\eqref{eq:x_update} is not differentiable as given, due to the $|\cdot|$ operator. Thus, we smoothen $|{\bf x}|$ using:
\begin{equation}\label{eq:Smoothing}
    |\mathbf{x}| \approx \sqrt{\mathbf{x}\odot\mathbf{x}^{\mathrm{H}}+\mathbf{\epsilon}}
\end{equation}
where $\mathbf{\epsilon}>0$, and $\odot$ is the Hadamard product. We refer to this approach as the quadratic smoothing (Quad-S) in the remainder of the paper. Beyond this strategy, more sophisticated smoothing formulations are investigated in an ablation study (\Cref{sec:AblationNetwork}). Without loss of generality and for ease of notation, we continue our formulation with Quad-S. We define: 
\begin{align}
    Q({\bf x}) \triangleq \|\mathbf{y}_\Omega &- \mathbf{E}_\Omega \mathbf{x}\|_2^2
            \: \:+ \mu \|\mathbf{x} - \mathbf{v}^{\mathrm{(i)}} + \mathbf{u}^{\mathrm{(i-1)}}\|_2^2 \nonumber \\
            &+\lambda \|\mathbf{r}_\Delta - \sqrt{(\mathbf{E}_\Delta \mathbf{x})\odot (\mathbf{E}_\Delta \mathbf{x})^{\mathrm{H}}+\mathbf{\epsilon}}\|_2^2 ,
\end{align}
with the derivative: 
\begin{align}\label{eq:DF_smooth}
    \nabla_{\bar{\mathbf{x}}} & Q(\mathbf{x}) = \: \mathbf{E}_\Omega^\mathrm{H}(\mathbf{E}_\Omega \mathbf{x} - \mathbf{y}_\Omega) + \mu (\mathbf{x} - \mathbf{v}^{\mathrm{(i)}} + \mathbf{u}^{\mathrm{(i-1)}}) \nonumber \\
    &+ \lambda \Bigg( \mathbf{E}_\Delta^\mathrm{H}\big(- \frac{\mathbf{r}_\Delta \odot (\mathbf{E}_\Delta \mathbf{x})}{\sqrt{(\mathbf{E}_\Delta \mathbf{x})\odot(\mathbf{E}_\Delta \mathbf{x})^{\mathrm{H}}+\mathbf{\epsilon}}} + \mathbf{E}_\Delta\mathbf{x} \big)\Bigg).
\end{align}
This can be used for a simple gradient descent (GD) as:
\begin{equation}\label{eq:simpleGD}
    \mathbf{x}_{\mathrm{j}}^{\mathrm{k}} = \mathbf{x}_{\mathrm{j}}^{\mathrm{k-1}} - \eta_{\mathrm{j}}^{\mathrm{k}} \nabla_{\bar{\mathbf{x}}} Q(\mathbf{x})\Big|_{{\bf x} = \mathbf{x}_{\mathrm{j}}^{\mathrm{k-1}}} 
\end{equation}
where $\eta_{\mathrm{j}}^{\mathrm{k}}$ denotes the learnable step size corresponding to the $\mathrm{j}^\textrm{th}$ GD iteration within the $\mathrm{k}^\textrm{th}$ unrolled step. However, due to the non-convexity introduced by the magnitude fidelity term, directly optimizing \Cref{eq:x_update} using standard GD may lead to poor local minima. To improve stability within the DF subproblem, we employ Nesterov acceleration~\cite{nesterov1983method}, which has previously shown improved convergence in inverse imaging and ADMM-based optimization frameworks~\cite{goldstein2014fast,thorley2021nesterov}: 
\begin{align}
    \mathbf{m}_\mathrm{j}^\mathrm{k}
    &= \beta \mathbf{m}_\mathrm{j-1}^\mathrm{k}
    + \nabla_{\bar{\mathbf{x}}} Q(\mathbf{x})
    \Big|_{\mathbf{x}=\mathbf{x}_\mathrm{j-1}^\mathrm{k}-\beta \mathbf{m}_\mathrm{j-1}^\mathrm{k}},
    \label{eq:Nesterov_momentum}\\
    \mathbf{x}_\mathrm{j}^\mathrm{k}
    &= \mathbf{x}_\mathrm{j-1}^\mathrm{k}
    - \eta_\mathrm{j}^\mathrm{k}\mathbf{m}_\mathrm{j}^\mathrm{k},
    \label{eq:Nesterov_update}
\end{align}
where $\mathbf{m}_\mathrm{j}^\mathrm{k}$ is the momentum direction, $\beta$ is the momentum coefficient and $\eta_\mathrm{j}^\mathrm{k}$ is a learnable step size. Alternative momentum-based optimization strategies, including Polyak momentum~\cite{polyak1964some} and Adam optimization~\cite{kingma2014adam}, are further investigated in our ablation studies (\Cref{sec:AblationNetwork}). We note that the proposed magnitude-informed DF formulation is independent of the choice of the learned regularizer, and can therefore be incorporated into both spatial and spatiotemporal PD-DL architectures. An overview of the proposed ${\mathbb C}+\textrm{Mag}$ unrolled PD-DL framework is shown in \Cref{fig:pipeline}.

\section{Imaging Experiments \& Implementation Details}
\subsection{Imaging Experiments}
All datasets used in this study were acquired under the corresponding institutional review board (IRB) approvals and data-sharing agreements. A summary of the datasets and acquisition parameters is provided in Table~\ref{tab:datasets}.

\subsubsection{Segmented Breath-Hold Cine MRI}
Datasets from the OCMR database~\cite{chen2020ocmr} were used. Preprocessing was performed to ensure consistent spatial and temporal dimensions across subjects, including readout (RO) oversampling removal, zero-padding along phase-encode (PE), and temporal interpolation to 25 cardiac phases. Retrospective undersampling was performed using a time-interleaved shifted equidistant sampling pattern~\cite{kellman2001TSENSE} at $\mathrm{R}\in\{6,8,12,16\}$ without ACS lines. Auxiliary magnitude information was obtained from neighboring cardiac phases within the same subject, \emph{without} requiring additional scan time or protocol modifications.  
Note in this setup, full k-space coverage is achieved across $\mathrm{R}$ adjacent cardiac phases, allowing the $\mathrm{R}-1$ nearest phases to provide auxiliary magnitude information for all locations in $\Omega^{\mathrm{C}}$.

\begin{figure*}[!t]
  \centering
  \includegraphics[width=\linewidth]{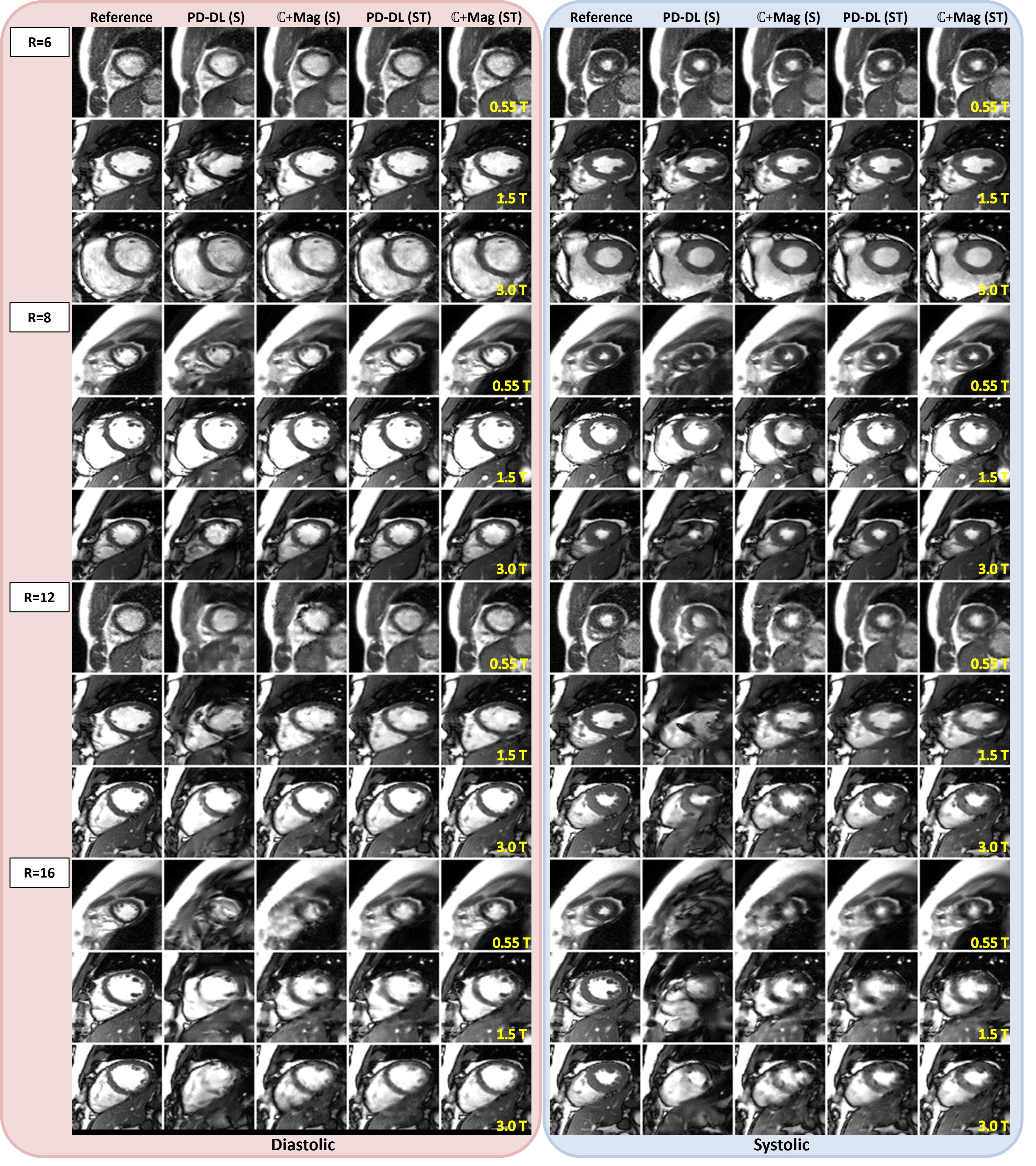}
  \vspace{-3ex}
  \caption{Representative cine MRI reconstructions at {\unboldmath $\mathrm{R} \in \{6,8,12,16\}$} from 0.55T, 1.5T, and 3.0T acquisitions, shown at diastolic (left) and systolic (right) cardiac phases. Both spatial-only (S) and spatiotemporal (ST) regularization is used for conventional PD-DL and the proposed ${\mathbb C} + $Mag formulations. Across acceleration factors, field strengths, and cardiac phases, ${\mathbb C} + $Mag yields improved anatomical fidelity and image sharpness relative to conventional PD-DL, with the ST regularizers further improving structural preservation and reducing reconstruction artifacts.}
  \vspace{-2ex}
  \label{fig:retro_cine_R6_R8}
\end{figure*}

\subsubsection{Phase-Contrast Flow MRI}
To evaluate the applicability of the proposed framework beyond cine imaging, axial aortic phase-contrast Flow2D MRI data from the CMRxRecon2025 challenge dataset\cite{cmrxrecon2025} 
were included. The dataset consists of single-slice acquisitions with two velocity encodes per timeframe. Preprocessing was performed similar to the segmented cine experiments. 
Retrospective undersampling and auxiliary magnitude extraction from neighboring temporal frames
were performed identically to the segmented cine experiments.

\subsubsection{Real-Time (RT) Cine MRI}
Free-breathing real-time cine MRI were acquired using balanced steady-state free precession (bSSFP) and gradient-echo (GRE) sequences.

\paragraph{bSSFP}
RT bSSFP cine data 
were obtained from the NIH Cardiac MRI Raw Data Repository hosted by the Intramural Research Program of the National Heart, Lung, and Blood Institute (NHLBI), in the short-axis view using a time-interleaved shifted equidistant undersampling trajectory. The data were acquired at $\mathrm{R}=4$. For our experiments, the data were further retrospectively undersampled to $\mathrm{R}=8$ by selecting every eighth PE line while additionally retaining the k-space center line for each timeframe~\cite{demirel2023high_EMBC,gulle2025_OVR}.

\paragraph{GRE}
RT GRE cine MRI data from 22 subjects 
were acquired locally at 3T in the short-axis view under an IRB-approved protocol, with an acceleration rate of $\mathrm{R}=8$.

\subsection{Implementation Details}
\subsubsection{Network Architecture} The proposed $\mathbb{C}+\text{Mag}$ PD-DL network was unrolled for $\mathrm{T}=10$ iterations. The DF unit was solved using Nesterov-accelerated GD as described in \Cref{eq:Nesterov_update}, which itself was unrolled for 10 iterations. All the DF parameters, including $\mu,\lambda, \gamma, \beta$ and step sizes $\eta_{\mathrm{j}}^{\mathrm{k}}$ were learned and unshared across the unrolls. The smoothing constant in~\Cref{eq:DF_smooth} was chosen as $\epsilon = 10^{-4}\cdot \| \mathbf{E}_\Omega^\mathrm{H}\mathbf{y}_\Omega\|_\infty^2$. 
The proximal step was implemented using both a spatial-only regularizer and a spatiotemporal regularizer, both based on
a recently proposed time-embedded U-Net (TE-UNet) architecture~\cite{junno2025_TE-MRI_NIPS} with \(\{96, 128, 160\}\) channels in the encoder and symmetric decoding layers. The spatial-only regularizer worked on each timeframe independently, thus with a 2-channel input/output. In contrast, the spatiotemporal regularizer worked on $\mathrm{R}$ neighboring timeframes, and thus had a $2\cdot \mathrm{R}$-channel input/output. During training, each temporal window batch was treated as an independent training sample and the loss was computed jointly over its $\mathrm{R}$ reconstructed frames, so overlapping windows required no additional handling. During testing, multiple estimates of the same cardiac phase arising from overlapping windows were averaged to obtain the final reconstruction. These two regularization settings were considered for complementary purposes: the spatial-only formulation isolates the contribution of the proposed auxiliary magnitude constraint from gains due to temporal modeling in the regularization, whereas the spatiotemporal formulation provides a stronger matched PD-DL baseline with access to the same neighboring-frame information and evaluates whether the proposed magnitude-informed DF remains beneficial when temporal correlations are already exploited by the learned regularizer. We emphasize that the proposed magnitude-informed DF formulation itself was the same across both regularizer setups, being applied to the given timeframe in the spatial-only setting and to each of the $\mathrm{R}$ timeframes in the spatiotemporal setting.

\begin{figure}[!t]
  \centering
  \includegraphics[width=0.95\linewidth]{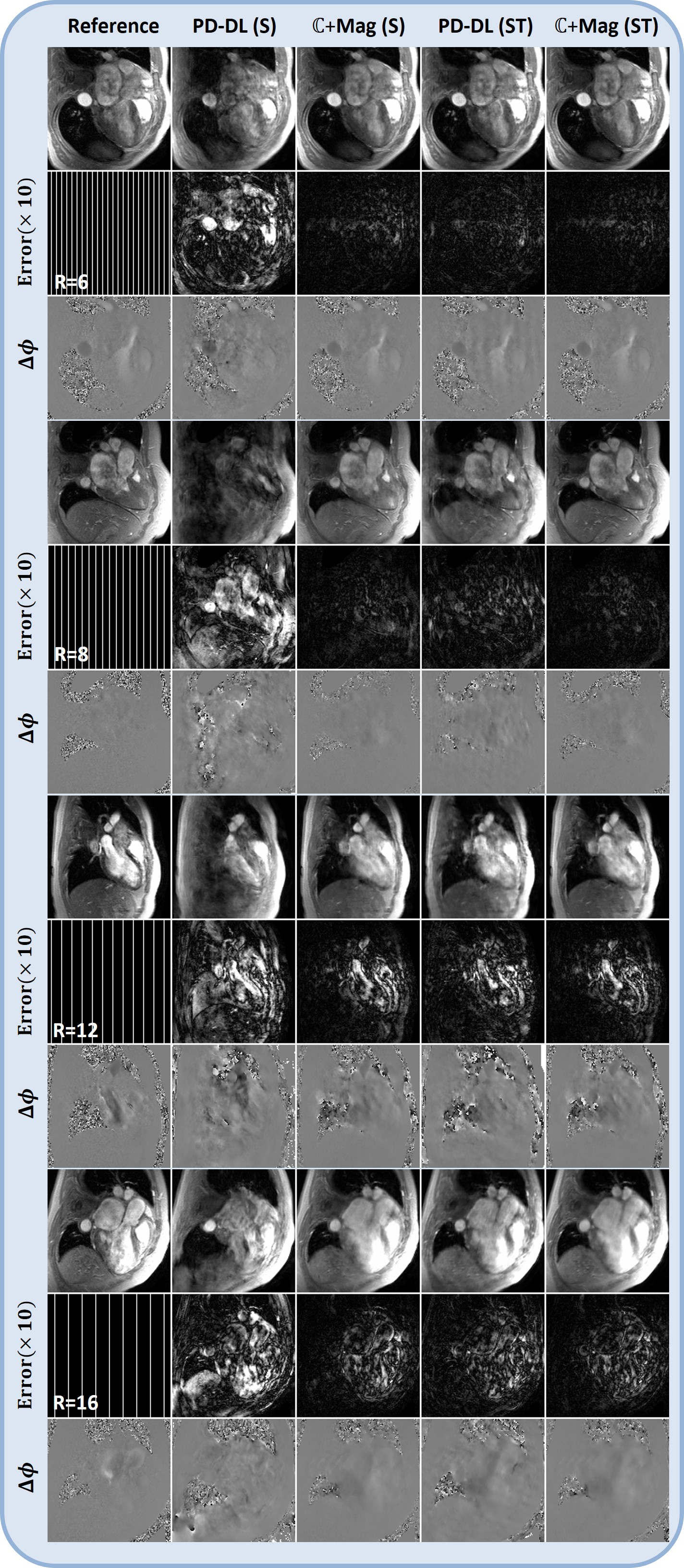}
  \caption{Representative reconstructions for phase-contrast Flow2D MRI at acceleration factors {\unboldmath $\mathrm{R}\in\{6,8,12,16\}$}. Here, (S) denotes spatial-only regularization and (ST) denotes spatiotemporal regularization. For each acceleration factor, the first row shows the reconstructed magnitude images, followed by the corresponding magnitude error maps ({\unboldmath $\times10$}) and phase-difference maps ({\unboldmath $\Delta\phi$}). Conventional PD-DL exhibits residual artifacts and phase inconsistencies, whereas the proposed $\mathbb{C}+\mathrm{Mag}$ formulation better suppresses these errors while preserving both magnitude and phase information across spatial-only (S) and spatiotemporal (ST) regularizer settings.}
  \label{fig:retro_r68_flow}
  \vspace{-2ex}
\end{figure}

\subsubsection{Auxiliary Magnitude Estimation}

For an acceleration factor $\mathrm{R}$, the shifted interleaved sampling pattern generates $\mathrm{R}$ distinct PE sampling patterns across cardiac phases, corresponding to $\mathrm{R}$ different sampling offsets. For each target phase $t$, we select one temporal frame from each residue class, choosing the available frame closest to $t$ for that sampling pattern. The target phase itself provides the complex-valued measurements on $\Omega$, while the remaining $\mathrm{R}-1$ selected phases provide magnitude information over the complementary locations $\Delta=\Omega^c$. The auxiliary magnitude estimate $\mathbf{r}_{\Delta}$ is then constructed by combining these complementary magnitude measurements.

\subsubsection{Comparison Methods}
Conventional PD-DL baselines employed the same TE-UNet proximal architectures, with both spatial-only and spatiotemporal regularization, and ADMM-based unrolling framework to ensure a fair comparison. Accordingly, the primary difference between methods was the underlying inverse problem formulation and corresponding DF unit. Specifically, conventional PD-DL methods solved the standard reconstruction objective in \eqref{eq:inverse_typical} using conjugate gradient (CG) for data fidelity~\cite{aggarwal2019MoDL}. Similarly, for the spatiotemporal PD-DL baseline, temporal information was incorporated directly into the learned regularizer by forming an $\mathrm{R}$-frame complex-valued input centered on the target phase. 

We emphasize that the proposed framework is not tied to a specific unrolled algorithm or proximal network architecture. Although ADMM unrolling is used in this work, the proposed complex-plus-magnitude formulation can in principle be incorporated into other unrolled reconstruction schemes, such as variable-splitting via quadratic penalty (VSQP) or proximal gradient descent-based (PGD) approaches.  
Similarly, the learned proximal operator can be replaced with alternative architectures.  
These design choices are further examined using the spatial-only $\mathbb{C}+\text{Mag}$ formulation in the ablation studies presented in \Cref{sec:AblationNetwork}. 

\subsubsection{Training} Supervised training was performed for retrospectively undersampled data \ie, segmented cine, and phase-contrast flow MRI), using SENSE-1 coil-combined images~\cite{knoll2020deep-survey} as reference data. All slices and timeframes from the training subjects were used for training. A normalized $\ell_1$-$\ell_2$ loss function was employed~\cite{saberi2024EMBC} and optimized using Adam with a learning rate of \(5\times 10^{-4}\). All models were trained for 100 epochs. 10\% of the training data was reserved for validation, hyperparameter fine-tuning, and learning rate scheduling. 

Multi-mask SSDU~\cite{yaman2022mmssdu} was employed for prospectively undersampled real-time GRE and bSSFP acquisitions. Similar training settings were used, except that the models were trained with learning rate of \(2\times 10^{-4}\), and three SSDU masks. 

\input{Tables/R1/quant_table}

\subsection{Evaluation}
\subsubsection{Standard Quantitative Metrics}  For datasets with available reference images (\ie, segmented cine and phase-contrast flow), reconstruction fidelity was quantitatively assessed using peak signal-to-noise ratio (PSNR) and structural similarity index measure (SSIM). 

\subsubsection{Cardiac Function Analysis} Cardiac function analysis was conducted on the segmented breath-hold cine, and prospective RT bSSFP and GRE datasets using Segment CMR (Medviso AB, Lund, Sweden). We note only four test subjects for segmented cine data had multiple slice acquisitions, and only these were used for quantification since multiple slices are required.
Endocardial and epicardial contours were delineated at end-diastole and end-systole to derive end-diastolic volume (EDV), end-systolic volume (ESV), stroke volume (SV), ejection fraction (EF), end-diastolic mass (EDM), and end-systolic mass (ESM). For the segmented cine datasets, measurements obtained from the reconstructed images were compared against those derived from the fully sampled reference acquisitions. For the prospective RT bSSFP datasets, the clinically used $\mathrm{R}=4$ tGRAPPA reconstruction~\cite{breuer2005dynamic} served as the reference standard. For the prospective GRE evaluation, six scans with matched breath-hold cine acquisitions were used, where the corresponding breath-held images served as the reference for volumetric analysis. Statistical differences between measurements were assessed using a paired t-test, with $P < 0.05$ considered statistically significant. All reconstruction models used for these cardiac function analyses employed the spatial-only regularization setting.

\subsubsection{Flow Analysis} 

For phase-contrast flow data, flow measurements were obtained from the descending aorta, including net flow volume (NetV), forward flow volume (FV), backward flow volume (BV), regurgitant fraction (RF), and cardiac output (CO), and were compared with the corresponding measurements from the reference data~\cite{heiberg2010design}. Additionally, we performed quantitative flow analysis, and computed the circular phase-difference error between the two Flow2D velocity encodes. For each cardiac frame, the phase difference was obtained as
$\Delta\phi=\angle\left(\mathbf{x}_2\odot\mathbf{x}_1^{*}\right)$, where $\mathbf{x}_1$ and $\mathbf{x}_2$ denote the two complex-valued velocity-encoded images. This was computed for both the reconstructed and reference data. To handle phase wrapping issues between the two different phase maps\cite{ong2018general, saberi2026umpire}, the wrapped phase error was defined as 
\[
\mathrm{e}_{\phi}=
\angle\left\{
\exp\left[j\left(\Delta\phi_{\mathrm{rec}}-\Delta\phi_{\mathrm{ref}}\right)\right]
\right\},
\]
and the mean absolute phase error (MAPE)$_{\Delta\phi}$ was evaluated over the foreground region as
\[
\text{MAPE}_{\Delta\phi}
=
\frac{1}{N}\sum_{i=1}^{N}\left|\mathrm{e}_{\phi,i}\right|.
\] 
Quantitative phase and flow analyses were performed on five subjects at $\mathrm{R}\in\{6,8\}$.
The quantitative flow measurements were performed using the spatial-only reconstructions at $\mathrm{R}\in\{6,8\}$, whereas the phase-error analysis was evaluated for all spatial-only and spatiotemporal models at $\mathrm{R}\in\{6,8,12,16\}$.

\subsubsection{Qualitative Expert Evaluation}

Qualitative image quality assessment was performed by an experienced cardiologist using a 4-point Likert scale~\cite{yaman2021_3D-LGE} for segmented breath-hold cine and 2D flow acquisitions. We note that only the four test subjects with multiple slices were used for the former dataset. In this blinded evaluation, the reader was blinded to the reconstruction method. Reconstructions were evaluated based on perceived SNR, blurring, aliasing artifacts, and overall image quality~\cite{hammernik2018VarNet}. For perceived SNR and overall image quality, scores were assigned as 1 (excellent), 2 (good), 3 (fair), and 4 (poor). Blurring was scored as 1 (no blurring), 2 (mild blurring), 3 (moderate blurring), and 4 (severe blurring). For aliasing artifacts, the scores rated as 1 (none), 2 (mild), 3 (moderate), and 4 (severe). The Wilcoxon signed-rank test was used to evaluate the reader scores, with a significance level of $P< 0.05$.
\newpage
\section{Results}
\subsection{Retrospectively Accelerated Datasets}
For these datasets, where fully-sampled k-space is available, we compare conventional PD-DL with two versions of the proposed $\mathbb{C}+\text{Mag}$ unrolled network. The first uses magnitude information estimated from adjacent cardiac phases, as proposed, while the second corresponds to an oracle setting using ground-truth magnitude information in \Cref{eq:inverse} directly. The oracle case is included to illustrate the upper performance bound for our framework, and quantify the gap between the proposed magnitude estimation strategy that does not require any additional acquisition time and the ideal scenario.

\begin{figure}[t]
  \centering
  \includegraphics[width=\linewidth]{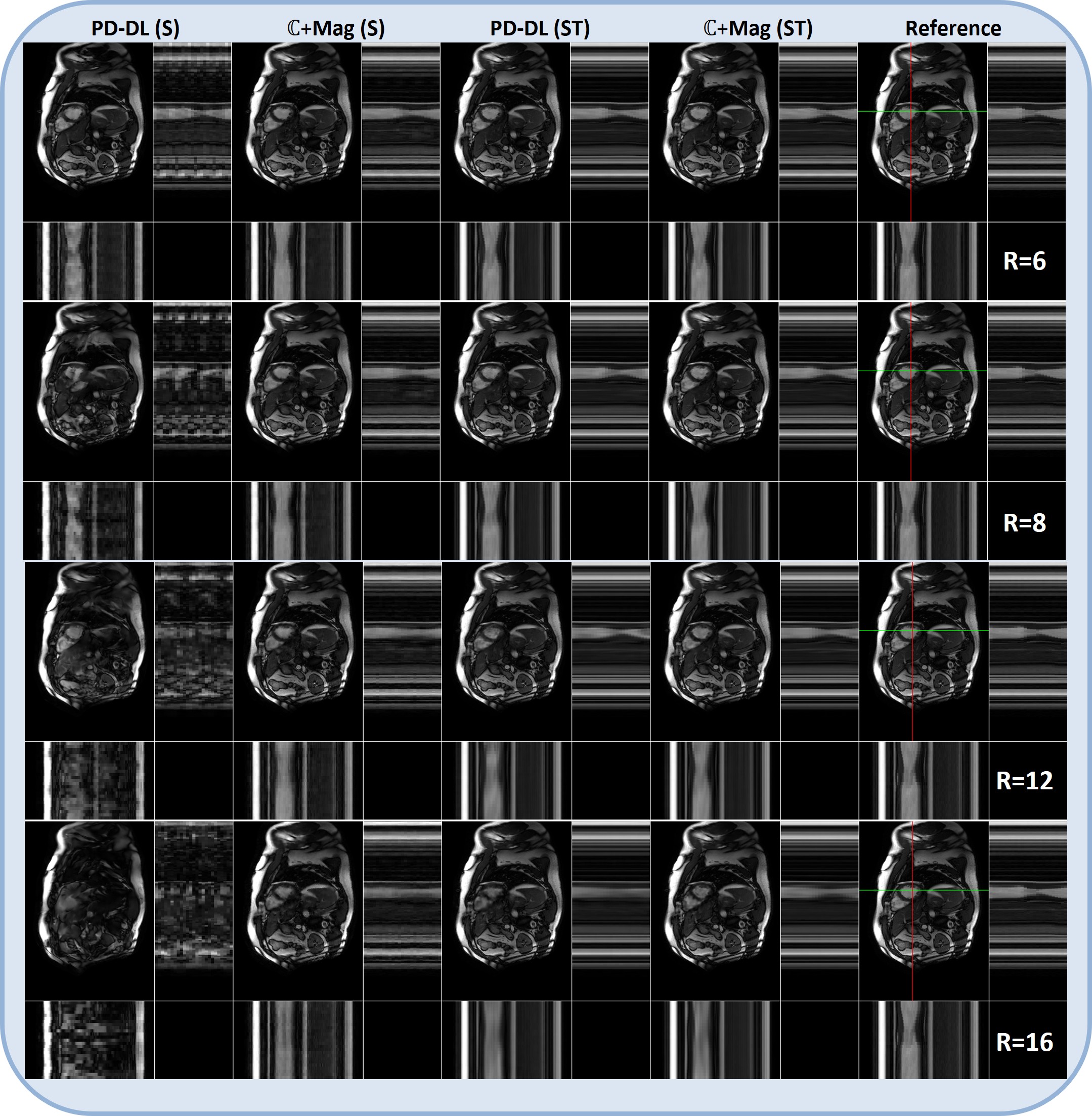}
  \caption{Spatiotemporal fidelity comparison for segmented cine MRI at $\mathrm{R}\in{6,8,12,16}$ with both spatial-only (S) and spatiotemporal (ST) regularizers, using conventional PD-DL and proposed ${\mathbb C} + $Mag. For each method, the reconstructed image is shown together with the corresponding x-t and y-t temporal profiles; the green and red lines in the reference indicate the selected profile locations. The temporal dimension is enlarged by $5\times$ for visualization.}
  \label{fig:xy_t}
\end{figure}

\subsubsection{Segmented Breath-Hold Cine MR}
\Cref{fig:retro_cine_R6_R8} shows representative reconstructions of systolic and diastolic cardiac phases at $\mathrm{R}\in\{6,8,12,16\}$. In the spatial-only regularizer setting, conventional PD-DL increasingly struggles to suppress undersampling artifacts at higher acceleration rates, leading to degraded visualization of cardiac structures. In contrast, the proposed $\mathbb{C}+\text{Mag}$ PD-DL substantially reduces these artifacts while preserving cardiac anatomy. In the spatiotemporal {regularizer} setting, incorporating neighboring cardiac phases into the learned proximal operator considerably improves the conventional PD-DL baseline. This is expected, as the spatiotemporal model has improved representation capacity to capture more complex cardiac dynamics. Importantly, the proposed $\mathbb{C}+\mathrm{Mag}$ framework does not aim to modify the regularizer itself, but instead improves the DF formulation. As such, the gains from spatiotemporal regularization and magnitude-informed DF are complementary, and integrating the proposed magnitude-informed DF with this spatiotemporal regularization provides additional improvements at $\mathrm{R}\in\{12,16\}$ over conventional PD-DL.
\Cref{fig:xy_t} further illustrates these observations through the corresponding x-t and y-t temporal profiles, showing improved preservation of temporal structures with spatiotemporal regularization and additional benefits from the proposed $\mathbb{C}+\mathrm{Mag}$ formulation at higher acceleration rates.
\Cref{tab:combined_results} summarizes the population-averaged quantitative results on the test set, which are consistent with these visual observations. We note that while the proposed $\mathbb{C}+\text{Mag}$ framework still visibly and quantitatively improves on conventional PD-DL in the spatiotemporal regularizer setup, the net gains are milder compared to the spatial-only setup, due to the stronger PD-DL baseline in the former case.

\subsubsection{Phase-Contrast Flow MRI} 
\Cref{fig:retro_r68_flow} show representative reconstruction results for 2D phase-contrast flow MRI at $\mathrm{R}\in\{6,8,12,16\}$. In the spatial-only setting, conventional PD-DL exhibits residual artifacts and phase inconsistencies, leading to degraded magnitude reconstruction and inaccurate phase difference maps in some regions. In contrast, the proposed $\mathbb{C}+\text{Mag}$ PD-DL effectively suppresses these artifacts while preserving both magnitude and phase information. The spatiotemporal regularizer formulation further improves reconstruction quality for Flow2D for both PD-DL and $\mathbb{C}+\mathrm{Mag}$ relative to the spatial-only setting. Incorporating the proposed $\mathbb{C}+\mathrm{Mag}$ DF within this stronger spatiotemporal model provides additional gains, demonstrating that the magnitude-informed constraint remains beneficial even when temporal information is already utilized by the regularizer. \Cref{tab:combined_results} summarizes the average population metrics on the test set. Consistent with the visual results, the proposed $\mathbb{C}+\text{Mag}$ PD-DL provides substantial quantitative improvements over conventional PD-DL in the spatial-only regularizer setting across all acceleration rates. In the spatiotemporal regularizer setting, the stronger conventional PD-DL baseline is further improved by incorporating the proposed $\mathbb{C}+\text{Mag}$. The improvement is most pronounced particularly at $\mathrm{R}=8$ and $12$, while the relative gain becomes smaller at $\mathrm{R}=16$, although $\mathbb{C}+\text{Mag}$ continues to outperform the corresponding spatiotemporal PD-DL baseline. 

\Cref{tab:flow_mape} summarizes mean $\text{MAPE}_{\Delta\phi}$ for the evaluated methods. The proposed $\mathbb{C}+\text{Mag}$ formulation consistently reduces the mean absolute phase error compared with conventional PD-DL across all acceleration rates. This improvement is particularly pronounced for the spatial-only models. For the spatiotemporal regularizers, the phase error is reduced substantially for both PD-DL and $\mathbb{C}+\text{Mag}$. Within this stronger spatiotemporal setting, $\mathbb{C}+\text{Mag}$ still consistently achieves lower phase error than the corresponding PD-DL baseline.

\input{Tables/R1/MAPE}

\input{Tables/R1/volumetric}

\subsection{Prospectively Undersampled Acquisitions}
We next investigate our method in prospectively undesampled real-time cine acquisitions. Unlike the retrospective setting, fully sampled reference data is not available in this setup, and therefore the oracle $\mathbb{C}+\text{Mag}$ experiment using ground-truth magnitude information cannot be performed.

\begin{figure}[t]
  \centering
  \includegraphics[width=\linewidth]{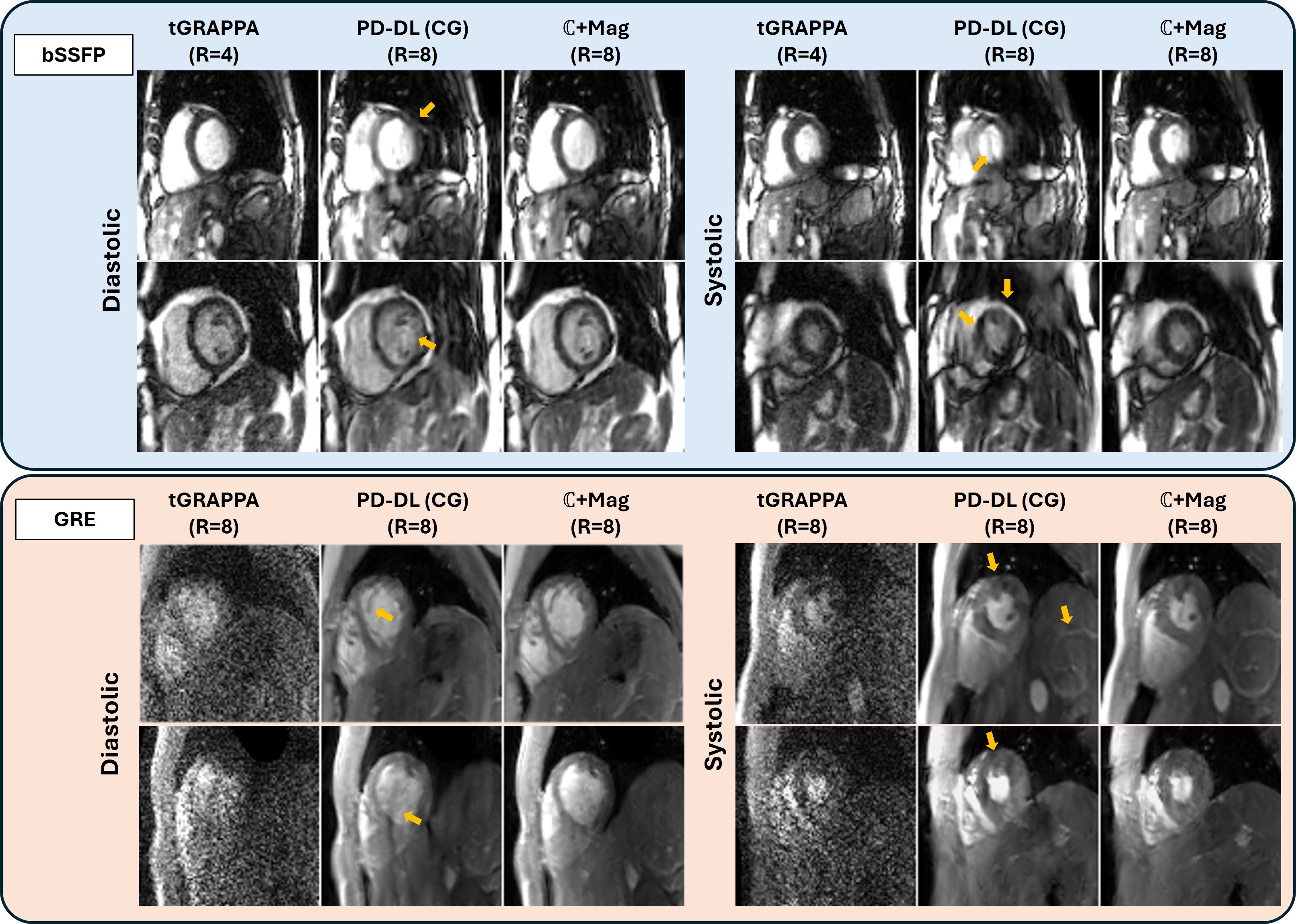}
  \caption{Reconstruction examples from prospectively undersampled bSSFP (top) and GRE (bottom) real-time cine data at {\unboldmath $\mathrm{R}=8$}. tGRAPPA at the clinical acquisition (\unboldmath$\mathrm{R}=4$) is used as the baseline for the bSSFP dataset. For the GRE dataset, the data were prospectively acquired at \unboldmath$\mathrm{R}=8$ and there is no \unboldmath$\mathrm{R}=4$ baseline. Standard tGRAPPA reconstruction and conventional PD-DL methods at \unboldmath$\mathrm{R}=8$ exhibit residual artifacts and loss of anatomical detail (yellow arrows), both in systolic and diastolic phases. The proposed $\mathbb{C}+\text{Mag}$ method preserves cardiac structures and suppresses reconstruction artifacts.}
  \label{fig:GRE_R8}
\end{figure}

\subsubsection{Real-Time Cine MR (bSSFP)}
\Cref{fig:GRE_R8} shows representative reconstructions from $\mathrm{R}=8$ real-time bSSFP cine data. While the clinically used tGRAPPA reconstruction at $\mathrm{R}=4$ provides a good image quality, conventional PD-DL methods at $\mathrm{R}=8$ exhibit noticeable residual artifacts and loss of anatomical detail. In contrast, the proposed $\mathbb{C}+\text{Mag}$ method that uses magnitude-only information from nearby time-frames, better preserves overall structures and suppresses reconstruction artifacts. Notably, despite the substantially higher acceleration factor, the proposed reconstruction at $\mathrm{R}=8$ achieves image quality comparable to the clinically used tGRAPPA reconstruction at $\mathrm{R}=4$, using spatial-only regularization.

\subsubsection{Real-Time Cine MR (GRE)}
\Cref{fig:GRE_R8} shows example reconstructions for a $\mathrm{R}=8$ acquisition. At this high acceleration rate, tGRAPPA exhibits substantial artifacts, as expected. Conventional PD-DL methods improve on tGRAPPA, but still suffer from residual artifacts and loss of anatomical detail in both systolic and diastolic phases. In contrast, the proposed $\mathbb{C}+\text{Mag}$ reconstruction effectively preserves cardiac structures while suppressing noise and residual artifacts.

\begin{figure}[!b]
  \centering
  \includegraphics[width=\linewidth]{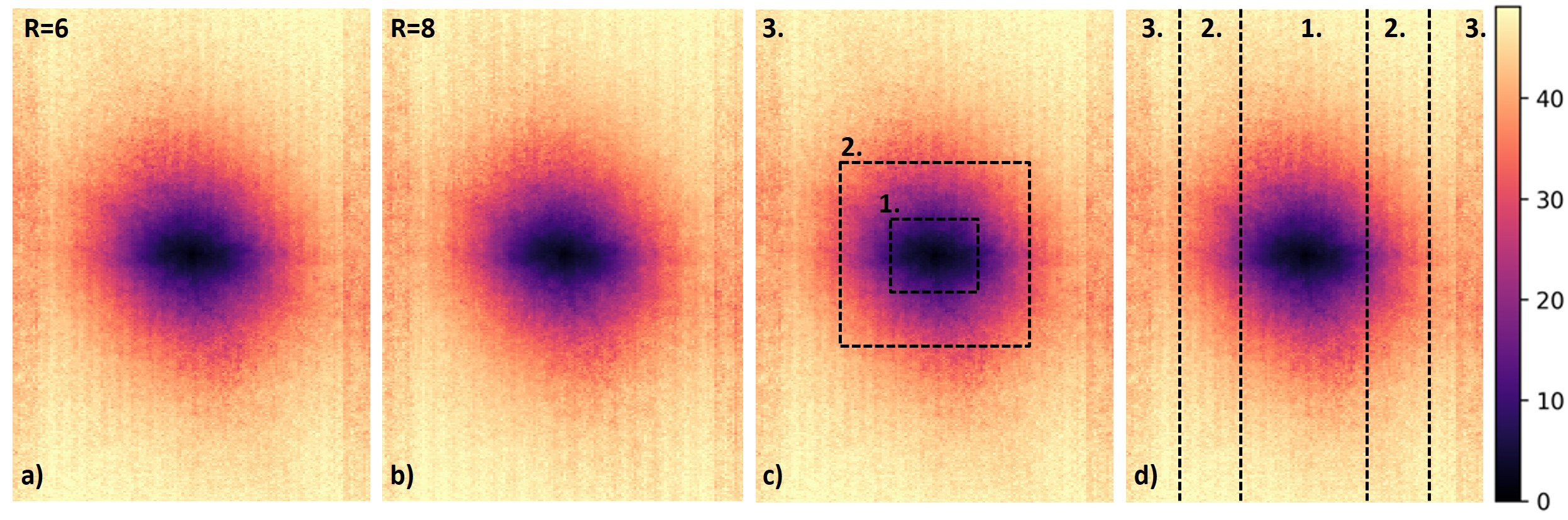}
  \caption{Relative error (RE) and spatial-frequency ablation for segmented cine. (a,b) Magnitude-estimation RE maps at $\mathrm{R}\in\{6,8\}$, respectively, showing lower RE near the k-space center and increasing RE toward higher spatial frequencies. Two complementary ablation partitions for experiments: (c) uses 2D spatial-frequency regions whose boundaries closely follow the observed relative-error distribution; (d) uses equal-width PE-direction regions that more closely reflect practical undersampling. Regions 1, 2, and 3 correspond to low-, mid-, and high-RE regions, respectively.}
  \label{fig:ablation_freq}
  \vspace{-2ex}
\end{figure}

\subsection{Quantitative Cardiac Function and Flow Analysis}
\Cref{tab:lv_analysis_all} summarizes the quantitative cardiac function and flow analyses. For cardiac function assessment, the proposed $\mathbb{C}+\text{Mag}$ reconstruction at $\mathrm{R}=8$ showed close agreement with the corresponding reference or baseline acquisitions across segmented breath-hold cine, real-time bSSFP, and real-time GRE datasets. The resulting volumetric and myocardial mass measurements were comparable to the reference values, with no statistically significant differences across the evaluated cardiac function metrics ($P>0.05$). Cardiac function analysis was performed at $\mathrm{R}=8$, representing the more challenging acceleration setting considered for these prospective and retrospective cine evaluations.

\input{Tables/ablations}
\input{Tables/R1/ablation_R1}

Similarly, for flow analysis,  the velocity-encoding phase was well preserved, and the resulting quantitative flow measurements remained closely aligned with the reference at both acceleration rates $\mathrm{R}\in\{6,8\}$. The derived NetV, FV, RF, and CO measurements showed close agreement with the reference values, with no statistically significant differences at either acceleration rate ($P>0.05$).

\subsection{Expert Cardiologist Readings}
\begin{figure*}[!t]
  \centering
  \includegraphics[width=0.88\linewidth]{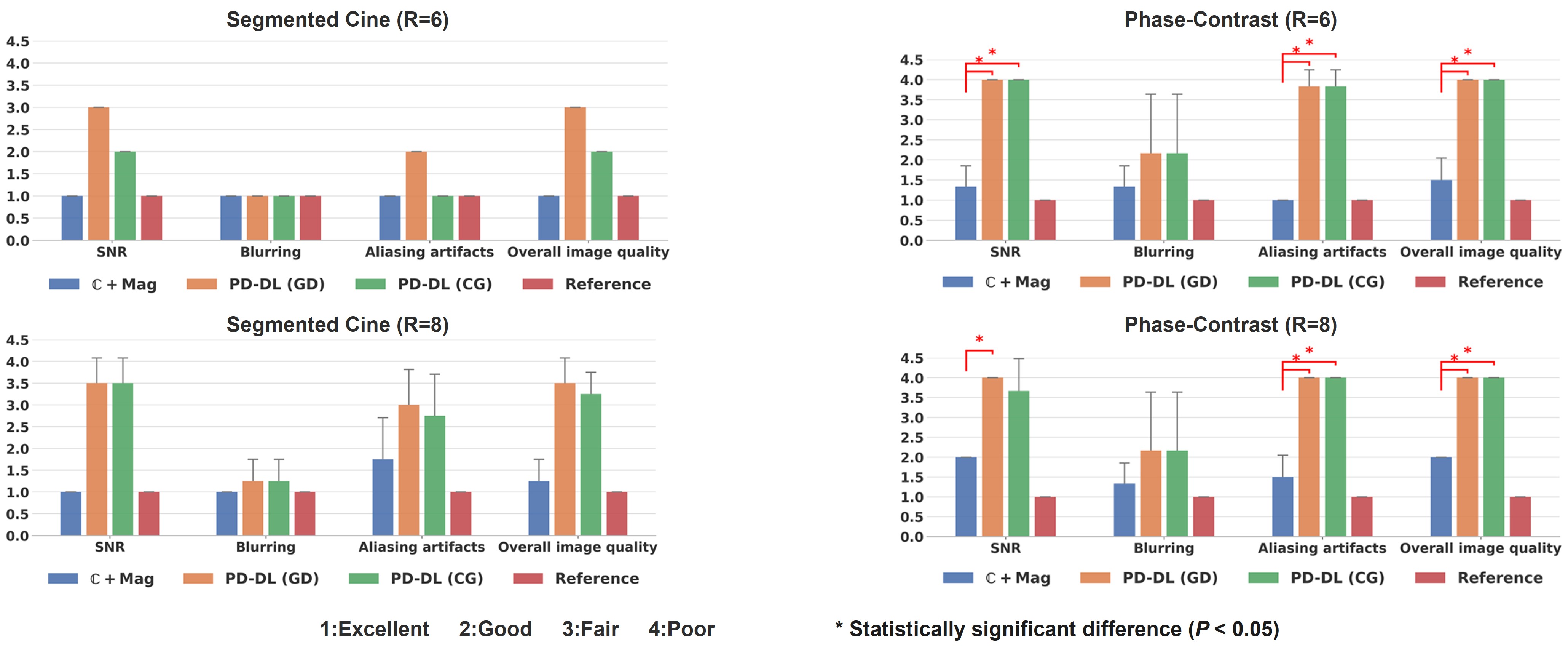}
  \caption{Results of the blinded expert reader evaluation for segmented cine and phase-contrast flow acquisitions at {\unboldmath \(\mathrm{R}=6\)} and {\unboldmath \(\mathrm{R}=8\)}. Bar plots show the mean reader scores and standard deviations for perceived SNR, blurring, aliasing artifacts, and overall image quality, where lower scores indicate better image quality. Statistical significance was assessed using Wilcoxon signed rank test. Across both datasets, the proposed \(\mathbb{C}+\text{Mag}\) reconstruction achieved scores comparable to the reference standard, while consistently outperforming conventional PD-DL reconstructions.}
  \label{fig:radio_assess}
\end{figure*}

\Cref{fig:radio_assess} summarizes the results of the blinded expert reader evaluation for cine and flow acquisitions. Across all evaluation criteria, the proposed method consistently received scores that were comparable to the corresponding reference acquisitions, while outperforming conventional PD-DL reconstructions substantially. In cine imaging, the proposed method demonstrated improved depiction of cardiac structures and reduced reconstruction artifacts, resulting in higher overall image quality, matching that of the fully-sampled reference. We note that the statistical power in this case is limited due to the small number of multi-slice cases scored. Similar trends were observed for phase-contrast flow imaging, where the proposed reconstructions provided superior visualization of vascular structures and flow-related features compared with conventional PD-DL methods. Notably, the reader scores obtained by the proposed method remained close to those of the reference images despite the substantially higher acceleration factor. These findings indicate that the proposed reconstruction preserves clinically relevant image features and enables accelerated cardiac MRI without compromising clinical image quality.

\subsection{Ablation Studies: Network Design Choices}\label{sec:AblationNetwork}

Our first set of ablation studies investigates the design choices of the unrolled reconstruction framework, including:
i) The network architecture for the proximal operator 
to show that the improvements of our framework is agnostic to the specifics of such network architectures, ii) Momentum-based GD schemes for solving the DF subproblem,  iii) Smoothing operators for the $|\cdot|$ term, and iv) Unrolling strategies. These experiments were performed at $\mathrm{R}=8$ with spatial only regularizers, and the corresponding population metrics on the full test set are reported in~\Cref{tab:ablation}. 

\subsubsection{Network architecture for proximal operator}
We utilize the ResNet architecture in~\cite{yaman2020SSDU, demirel2023SIIM, saberi2026ICML} to solve \Cref{eq:v_update} instead of the baseline time-embedded U-Net~\cite{junno2025_TE-MRI_NIPS}, and also 
investigate the impact of sharing the DF learnable parameters, while keeping the regularizer parameters shared. These experiments utilize a simple GD for solving~\Cref{eq:x_update}. 
The results indicate that the TE-UNet with unshared DF parameters achieves the best performance, confirming the theoretical findings in~\cite{junno2025_TE-MRI_NIPS}. It is worth noting that even the simpler setup of ResNet with shared parameters achieves substantial improvements over conventional PD-DL methods, highlighting the benefits of our inverse problem formulation in~\Cref{eq:inverse}, irrespective of architectural optimization.

\subsubsection{Solving the data fidelity sub-problem} With $\textrm{prox}_{\cal R}(\cdot)$ architecture fixed, we investigate different three momentum-based GD schemes for solving the DF subproblem in \Cref{eq:x_update}, as discussed in \Cref{sec:Method}. Among these approaches, Nesterov-GD (denoted as Nesterov) achieves the best performance. 

\subsubsection{Smoothing for the magnitude term.} We further investigate the effect of the smoothing formulation for \Cref{eq:Smoothing}. In addition to the proposed simple quadratic smoothing (Quad-S), we evaluate alternative smooth approximations of the absolute value operator, including Smooth-$\ell_1$ (Huber) smoothing~\cite{huber1964robust}, Pseudo-Huber (PH-S)~\cite{charbonnier1997deterministic}, and Log-Exp (Log-S) smoothing~\cite{nesterov2005smooth}. The corresponding hyperparameters were fine-tuned on a small validation subset for maximal performance ($\delta=1$ for Smooth-$\ell_1$ and Pseudo-Huber smoothing, and $\alpha=20$ for Log-Exp smoothing). However, none of these alternatives outperformed the proposed Quad-S formulation, which consistently achieved the best performance, despite its simplicity, on the full test set.

\subsubsection{Unrolling strategies} Finally, to highlight that our formulations are agnostic to the specifics of the algorithm unrolling strategy, we also investigate different unrolling strategies. As reported in~\Cref{tab:ablation}, the ADMM-based formulation achieves the best performance, outperforming both variable splitting with quadratic penalty (VSQP)~\cite{aggarwal2019MoDL} and proximal gradient descent (PGD)~\cite{mardani2018neural, hosseini2020dense}. While VSQP maintains competitive quantitative performance, PGD exhibits a substantial degradation in performance. To assess whether this degradation is primarily due to the smaller number of DF updates, we additionally evaluate PGD with 10 inner DF iterations per unroll, which substantially improves performance; however, despite using the same number of DF steps, it remains below both VSQP and ADMM. These results indicate that the performance differences cannot be attributed solely to the amount of iterative computation and further support the benefit of the ADMM-based formulation.
\input{Tables/R1/ablation_compute}
Consistently, the computational analysis in~\Cref{tab:compute_cost} shows that PGD with 10 inner iterations, VSQP, and ADMM have comparable computational budgets, runtimes, memory requirements, and identical model capacities. Their per-sample inference times were approximately 175--189 ms, while the corresponding training times were approximately 534--570 ms per sample. Therefore, the remaining performance differences are primarily attributable to the unrolling strategy itself, indicating that the choice of optimization formulation plays a role beyond simply increasing the amount of DF computation.

\subsection{Ablation Studies: Magnitude Estimation}\label{sec:AblationMagEst}

Our second set of ablation studies investigates the role of the auxiliary magnitude information by studying: i) the contribution of different spatial-frequency regions, guided by the relative-error distribution, at $\mathrm{R}\in\{6,8\}$; and
ii) the number, distribution, and temporal distance of the auxiliary cardiac phases used for magnitude estimation at $\mathrm{R}=8$.
The corresponding population metrics on the full test set are reported in~\Cref{tab:ablation_freq_temporal}.

\subsubsection{Spatial-frequency analysis}\label{sec:ablation_freq_relative} 
We first set out to study the behavior of c-sim across different frequency regions. We partitioned $\Delta$ into low-, mid-, and high-frequency bands based on normalized radial frequency and evaluate c-sim within each band. For each normalized k-space coordinate ${\bf k} = (k_x,k_y) \in [-1,1]^2,$ we computed the normalized radial frequency as $\rho=\sqrt{(\mathrm{k}_\mathrm{x}^2+\mathrm{k}_\mathrm{y}^2)/2} \in [0,1]$. The low-, mid-, and high-frequency regions were defined as $\rho<1/3$, $1/3\leq\rho<2/3$, and $\rho\geq2/3$, respectively. As expected, the c-sim was highest at low spatial frequencies with higher intensities, and gradually decreased toward higher frequencies (cine: 0.997, 0.881 and 0.730; flow: 0.995, 0.882 and 0.723 across low-, mid-, and high-frequency regions). Although these median values were similar, the phase-contrast data distributions were generally tighter, consistent with the greater acquisition uniformity of the single-center dataset.

This motivated us to further characterize the magnitude-estimation error as a function of the relative error (RE), instead of bulk alignment of intensities. For a k-space coordinate, ${\bf k}$, RE was defined as: \[
\text{RE}(\mathbf{k})
=
\frac{\left| \mathbf{r}(\mathbf{k})-\left|\mathbf{y}(\mathbf{k})\right| \right|}
{\left|\mathbf{y}(\mathbf{k})\right|},
\]where ${\bf y}$ is the true k-space value and ${\bf r}$ is the estimated magnitude from adjacent cardiac phases. The RE map across the k-space is depicted in \Cref{fig:ablation_freq}(a,b), and was obtained by first taking the median across coils at each ${\bf k}$, followed by the median across slices and cardiac phases within each subject, and across subjects. As expected, the relative error increased toward higher spatial frequencies, while lower frequency regions exhibit low relative error due to their higher intensities.

Accordingly, we designed two complementary experiments to evaluate the contribution of $\mathbf{r}_{\Delta}$ across these regions.  
The first \big(\Cref{fig:ablation_freq}(c)\big) was chosen based on 2D rectangular frequency regions with one central low-frequency region, one intermediate relative error region, and a third
high relative error region corresponding to outer k-space. The second set of regions followed a similar trend, but was selected to be partitioned only along the phase encode direction \big(\Cref{fig:ablation_freq}(d)\big) to align with physical undersampling constraints. These ablations were repeated with both estimated magnitudes and oracle magnitudes to get a full picture. 
As shown in \Cref{tab:ablation_freq_temporal}(a), for both regions and magnitude estimation methods, the highest gains in reconstruction 
quality is achieved when magnitude side information is used in the low-frequency region, whereas mid- and high-frequency regions provide smaller improvements. This is consistent with the higher relative error in magnitude per k-space point. We also highlight that running the proposed algorithm across all three regions (\ie all $\Delta$), still outperforms the low-frequency region only, even with the higher relative error in magnitudes in Regions 2 and 3.

\subsubsection{Temporal distance analysis} The utility of auxiliary magnitude information is governed by two factors: the amount and distribution of complementary information, and the temporal distance of the auxiliary cardiac phases from the target timeframe. We therefore evaluate these two aspects separately to characterize their respective contributions to reconstruction performance. 
First, we varied the number ($k$) and temporal distribution of auxiliary phases while retaining the same shifted interleaved sampling pattern~\cite{kellman2001TSENSE}. For a fixed number of auxiliary phases, the {Near}, {Intermediate}, and {Far} configurations progressively increase their temporal separation from the target phase (e.g., for $k=2$, we use $\{t-1, t+1\}$, $\{t-2, t+2\}$, and $\{t-3, t+3\}$, respectively), which effectively evaluates the amount and distribution of auxiliary phases. As shown in \Cref{tab:ablation_freq_temporal}(b)-i, performance improves with more auxiliary cardiac phases, and for a fixed number of cardiac phases with broader complementary k-space coverage. This suggests that increased coverage of $\Delta$ can outweigh the reduced reliability of more temporally distant phases. 

Second, we isolated the effect of temporal distance by shifting adjacent cardiac phases by $\mathrm{R}/2$ PE lines, producing two alternating sampling patterns that provide the same complementary locations from phases at different temporal distances. 
Specifically, for a target phase $t$, $\{t+1, t+3, t+5\}$ provide
magnitude information at the same complementary k-space location. We note that, in this setting, the auxiliary cardiac phases do not collectively provide full k-space coverage, which is not an inherent requirement of our method. As shown in \Cref{tab:ablation_freq_temporal}(b)-ii, the performance is highest when the magnitude information comes from the temporally closest cardiac phase
and gradually decreases with temporal distance, confirming greater reliability of temporally closer auxiliary cardiac phases. We note that the proposed \(\mathbb{C}+\text{Mag}\) outperforms the PD-DL baseline in all cases, even with information from more distant temporal frames. This also highlights a setting in which, even without full k-space coverage from auxiliary cardiac phases, the proposed method still outperforms conventional PD-DL.

\section{Limitation}
The proposed framework relies on the temporal consistency of the auxiliary k-space magnitude information and may therefore be affected by irregular cardiac motion.
In such settings, RT cine is preferred over segmented cine, since neighboring frames are acquired sequentially with snapshot imaging rather than being assembled across multiple heartbeats.
Consistently, our temporal-distance ablation shows gradually reduced performance with increasing temporal distance, which in effect would be similar to sudden changes due to cardiac motion, although all configurations remain above conventional PD-DL, still highlighting the potential benefits of our method for RT cine MRI, even in this challenging setup. 

In this work, we focused on steady-state dynamic MRI due to the stability of k-space magnitude across timeframes. The applicability of the proposed framework to other dynamic MRI applications depends on the temporal stability of the underlying signal. Extension to other steady-state applications in non-cardiac setups, such as carotid flow imaging, may be feasible when neighboring frames exhibit similar k-space magnitudes. However, extension to non-steady-state applications, such as perfusion or dynamic contrast-enhanced MRI, is less straightforward because substantial temporal signal changes may reduce the reliability of auxiliary magnitude information. In such settings, alternative strategies for identifying sufficiently consistent auxiliary measurements may be required.

\section{Discussion and Conclusion}
In this work, we proposed a PD-DL framework for incorporating auxiliary magnitude k-space information into MRI reconstruction. We identified steady-state dynamic MRI as a practical application of this framework, where informative auxiliary magnitude information can be obtained from neighboring time-frames without incurring additional acquisition cost. Experimental results on retrospectively undersampled cine, phase-contrast flow MRI data at $\mathrm{R}\in\{6,8\}$, and prospectively undersampled real-time cine MRI data, demonstrated substantial improvements over conventional PD-DL reconstruction methods in both image quality and phase preservation.

 Finally, we note that the focus of this work is to introduce a \emph{new inverse problem formulation} that enables incorporation of auxiliary magnitude information into PD-DL MRI reconstruction, in particular for dynamic MRI with no additional acquisition costs. Advances in network architectures, regularizers, and optimization strategies are complementary to our proposed framework and can be naturally integrated within it. Indeed, our ablation studies demonstrate that our proposed formulation can be synergistically combined with different optimization schemes and network architectures. Thus, further comparisons of different network architectures, including those that enable spatiotemporal regularization, are not critical for this work.

\section*{Acknowledgment}
We thank Dr. Peter Kellman for providing the real-time cine bSSFP data. The authors also acknowledge the Intramural Research Program of the NHLBI for supporting the NIH Open Source Cardiac MRI Raw Data Repository used in this study.

\bibliographystyle{IEEEtran}
\bibliography{main_clean}

\end{document}

%% file: Tables/acqusition_params.tex
\begin{table*}[!t]
\centering
\caption{Summary of imaging datasets and acquisition parameters used in this study.SD: Scan-dependent.} 
\label{tab:datasets}

\begin{adjustbox}{max width=\textwidth}
{\footnotesize
\renewcommand{\arraystretch}{0.9}
\setlength{\tabcolsep}{9.0pt}

\begin{tabular}{@{}p{3.5cm}cccc}
\toprule
 & Segmented Cine & Flow2D & Real-Time Cine (bSSFP) & Real-Time Cine (GRE) \\
\midrule
Sequence Type
& bSSFP
& Phase-Contrast Flow2D
& bSSFP
& GRE \\

Field Strength
& 0.55T / 1.5T / 3.0T
& 3.0T 
& 1.5T
& 3.0T \\

View
& Short-axis
& Axial aortic
& Short-axis
& Short-axis \\

{Subjects (Train/Test)}
& 46 / 14
& 20 / 8
& 10 / 3
& 10 / {12} \\

Matrix Size
& $288\times208$
& $224\times160$
& $160\times96$ 
& $176\times176$ \\

Field-of-view
& SD
& $340\times234\,\mathrm{mm}^2$
& $360\times270\,\mathrm{mm}^2$ 
& $300\times300\,\mathrm{mm}^2$ \\

Spatial Resolution
& $1.17-2.75 \: \mathrm{mm}$
& $1.77\times1.95\,\mathrm{mm}^2$
& $2.25\times2.25\,\mathrm{mm}^2$
& $1.70\times1.70\,\mathrm{mm}^2$ \\

Slice Thickness
& $6-8 \:\mathrm{mm}$
& --- 
& $8\,\mathrm{mm}$
& $5\,\mathrm{mm}$ \\

Acquisition Acceleration
& Fully-sampled
& Fully-sampled
& $\mathrm{R}=4$
& $\mathrm{R}=8$ \\

Evaluation Acceleration
& $\mathrm{R}\in\{6,8,12,16\}$
& $\mathrm{R}\in\{6,8,12,16\}$
& $\mathrm{R}=8$
& $\mathrm{R}=8$ \\

Partial Fourier
& ---
& ---
& $6/8$
& $6/8$ \\

Asymmetric Echo
& ---
& ---
& $20\%$
& $20\%$ \\

TR / TE (ms)
& SD 
& $37.1 / 2.5$
& $2.3 / 1.0$
& $4.7 / 2.4$ \\

\bottomrule
\end{tabular}
}
\end{adjustbox}
\vspace{-2ex}
\end{table*}

%% file: Tables/R1/quant_table.tex
\begin{table*}[!t]
\vspace{-2ex}
\caption{Quantitative comparison on retrospectively undersampled cine and phase-contrast Flow2D MRI data at $\mathrm{R}\in\{6,8,12,16\}$. The spatiotemporal proximal operators incorporate neighboring complex-valued frames along the channel dimension of the regularizer, leading to improved baseline performance. Note that \colorbox{gray!15}{$\mathbb{C}+\text{Mag}$ (Oracle)} is included to show the upper performance bound of our inverse problem, but is not practical for magnitude estimation.
}
\label{tab:combined_results}

\centering
\begin{adjustbox}{max width=\textwidth}
\begin{footnotesize}
\renewcommand{\arraystretch}{1.12}
\setlength{\tabcolsep}{4.0pt}

\begin{tabular}{@{}lcccccccccccccccc@{}}
\toprule

\multirow{4}{*}{Method}
& \multicolumn{8}{c}{Segmented Cine MRI}
& \multicolumn{8}{c}{Phase-Contrast Flow2D MRI} \\

\cmidrule(lr){2-9}
\cmidrule(lr){10-17}

& \multicolumn{2}{c}{$\mathrm{R=6}$}
& \multicolumn{2}{c}{$\mathrm{R=8}$}
& \multicolumn{2}{c}{$\mathrm{R=12}$}
& \multicolumn{2}{c}{$\mathrm{R=16}$}
& \multicolumn{2}{c}{$\mathrm{R=6}$}
& \multicolumn{2}{c}{$\mathrm{R=8}$}
& \multicolumn{2}{c}{$\mathrm{R=12}$}
& \multicolumn{2}{c}{$\mathrm{R=16}$} \\

\cmidrule(lr){2-3}
\cmidrule(lr){4-5}
\cmidrule(lr){6-7}
\cmidrule(lr){8-9}
\cmidrule(lr){10-11}
\cmidrule(lr){12-13}
\cmidrule(lr){14-15}
\cmidrule(lr){16-17}

& {\scriptsize PSNR$\uparrow$} & {\scriptsize SSIM$\uparrow$}
& {\scriptsize PSNR$\uparrow$} & {\scriptsize SSIM$\uparrow$}
& {\scriptsize PSNR$\uparrow$} & {\scriptsize SSIM$\uparrow$}
& {\scriptsize PSNR$\uparrow$} & {\scriptsize SSIM$\uparrow$}
& {\scriptsize PSNR$\uparrow$} & {\scriptsize SSIM$\uparrow$}
& {\scriptsize PSNR$\uparrow$} & {\scriptsize SSIM$\uparrow$}
& {\scriptsize PSNR$\uparrow$} & {\scriptsize SSIM$\uparrow$}
& {\scriptsize PSNR$\uparrow$} & {\scriptsize SSIM$\uparrow$} \\

\midrule

\multicolumn{17}{l}{\colorbox{orange!15}{Spatial-only}} \\[0.3ex]

PD-DL (CG)
& 34.67 & 0.927
& 29.93 & 0.849
& 26.85 & 0.757
& 25.61 & 0.715
& 26.51 & 0.786
& 24.21 & 0.690
& 23.25 & 0.649
& 23.00 & 0.626 \\

{\bf $\mathbb{C}$+Mag (Ours)}
& {\bf 40.69} & {\bf 0.974}
& {\bf 38.67} & {\bf 0.963}
& {\bf 35.92} & {\bf 0.940}
& {\bf 33.83} & {\bf 0.923}
& {\bf 34.67} & {\bf 0.933}
& {\bf 32.12} & {\bf 0.905}
& {\bf 29.50} & {\bf 0.856}
& {\bf 28.92} & {\bf 0.847} \\

\rowcolor{gray!15}
$\mathbb{C}$+Mag (Oracle)
& 42.45 & 0.980
& 41.15 & 0.975
& 38.45 & 0.958
& 36.67 & 0.947
& 40.87 & 0.976
& 37.59 & 0.962
& 33.84 & 0.925
& 31.70 & 0.883 \\

\midrule

\multicolumn{17}{l}{\colorbox{blue!15}{Spatiotemporal}} \\[0.3ex]

PD-DL (CG)
& {\bf 47.49} & {\bf 0.991}
& 45.36 & {\bf 0.988}
& 41.20 & 0.979
& 38.14 & 0.969
& 36.92 & 0.959
& 34.30 & 0.938
& 31.57 & 0.908
& 30.91 & 0.900 \\

{\bf $\mathbb{C}$+Mag (Ours)}
& 47.29 & {\bf 0.991}
& {\bf 45.40} & {\bf 0.988}
& {\bf 41.77} & {\bf 0.980}
& {\bf 38.99} & {\bf 0.972}
& {\bf 37.68} & {\bf 0.964}
& {\bf 35.30} & {\bf 0.947}
& {\bf 32.85} & {\bf 0.924}
& {\bf 31.75} & {\bf 0.905} \\

\rowcolor{gray!15}
$\mathbb{C}$+Mag (Oracle)
& 47.68 & 0.992
& 46.53 & 0.991
& 45.41 & 0.989
& 44.09 & 0.986
& 42.92 & 0.984
& 41.21 & 0.980
& 38.83 & 0.970
& 36.68 & 0.957 \\

\bottomrule
\end{tabular}

\end{footnotesize}
\end{adjustbox}
\end{table*}

%% file: Tables/R1/MAPE.tex
\begin{table}[!t]
\centering
\caption{Mean absolute phase error of the velocity-encoding phase difference, $\mathrm{MAPE}_{\Delta\phi}$ (rad), for retrospectively undersampled Flow2D MRI at $\mathrm{R}\in\{6,8,12,16\}$. Results are reported as mean $\pm$ standard deviation over subjects.}
\label{tab:flow_mape}
\renewcommand{\arraystretch}{1.12}
\setlength{\tabcolsep}{6pt}
\scriptsize

\begin{adjustbox}{max width=\columnwidth}
\begin{tabular}{lcccc}
\toprule
\textbf{Method} & $\mathrm{R=6}$ & $\mathrm{R=8}$ & $\mathrm{R=12}$ & $\mathrm{R=16}$ \\
\midrule

\multicolumn{5}{l}{\colorbox{orange!15}{Spatial-only}} \\[0.3ex]

PD-DL (CG)
& 0.112 $\pm$ 0.029
& 0.168 $\pm$ 0.052
& 0.186 $\pm$ 0.048
& 0.224 $\pm$ 0.045 \\

$\mathbb{C}+\text{Mag}$ (Ours)
& {\bf 0.072} $\pm$ {\bf 0.016}
& {\bf 0.088} $\pm$ {\bf 0.018}
& {\bf 0.120} $\pm$ {\bf 0.025}
& {\bf 0.119} $\pm$ {\bf 0.026} \\

\rowcolor{gray!15}
$\mathbb{C}+\text{Mag}$ (Oracle)
& 0.037 $\pm$ 0.011
& 0.051 $\pm$ 0.011
& 0.077 $\pm$ 0.014
& 0.104 $\pm$ 0.028 \\

\midrule

\multicolumn{5}{l}{\colorbox{blue!15}{Spatiotemporal}} \\[0.3ex]

PD-DL (CG)
& 0.050 $\pm$ 0.007
& 0.060 $\pm$ 0.008
& 0.073 $\pm$ 0.011
& 0.078 $\pm$ 0.012 \\

$\mathbb{C}+\text{Mag}$ (Ours)
& {\bf 0.049} $\pm$ {\bf 0.007}
& {\bf 0.058} $\pm$ {\bf 0.008}
& {\bf 0.070} $\pm$ {\bf 0.011}
& {\bf 0.077} $\pm$ {\bf 0.012} \\

\rowcolor{gray!15}
$\mathbb{C}+\text{Mag}$ (Oracle)
& 0.030 $\pm$ 0.006
& 0.034 $\pm$ 0.006
& 0.041 $\pm$ 0.007
& 0.049 $\pm$ 0.008 \\

\bottomrule
\end{tabular}
\end{adjustbox}
\end{table}

%% file: Tables/R1/volumetric.tex
\begin{table}[!b]
\centering
\caption{Cardiac function analysis for cine sequences and quantitative flow analysis for Flow2D.
EDV: end-diastolic volume; ESV: end-systolic volume; EF: ejection fraction;
SV: stroke volume; EDM: end-diastolic mass; ESM: end-systolic mass;
NetV: net flow volume; FV: forward flow volume; BV: backward flow volume;
RF: regurgitant fraction; CO: cardiac output.
Flow measurements were obtained from the descending aorta.
Results are given as mean, with the standard deviation in parentheses.
Differences between the proposed method and baseline were non-significant
({\unboldmath $P > 0.05$}) across all metrics.
}
\scriptsize
\label{tab:lv_analysis_all}

\renewcommand{\arraystretch}{1.15}
\setlength{\tabcolsep}{8pt}

\begin{adjustbox}{max width=\columnwidth}
\begin{tabular}{lcccccc}
\toprule

\multirow{3}{*}{\textbf{Metric}}
& \multicolumn{2}{c}{\textbf{Segmented BH Cine}}
& \multicolumn{2}{c}{\textbf{Real-Time Cine (bSSFP)}}
& \multicolumn{2}{c}{{\textbf{Real-Time Cine (GRE)}}}\\

\cmidrule(lr){2-3}
\cmidrule(lr){4-5}
\cmidrule(lr){6-7}

& \makecell{$\mathbb{C}+\text{Mag}$\\($\mathrm{R}=8$)}
& \makecell{Reference\\($\mathrm{R}=1$)}
& \makecell{$\mathbb{C}+\text{Mag}$\\($\mathrm{R}=8$)}
& \makecell{tGRAPPA\\($\mathrm{R}=4$)}
& \makecell{$\mathbb{C}+\text{Mag}$\\($\mathrm{R}=8$)}
& \makecell{Breath-\\Hold}
\\

\midrule

EDV (mL)
& 119.6 (26) & 121.8 (25)
& 98.7 (14) & 100.7 (13)
& 120.7 (20.7) & 121.3 (19) \\

ESV (mL)
& 48.3 (19) & 48.4 (18)
& 75.0 (30) & 78.3 (23)
& 36.7 (12.8) & 37.0 (13) \\

SV (mL)
& 71.3 (19) & 73.2 (19)
& 23.7 (16) & 22.3 (11)
& 84.0 (10.5) & 84.3 (8) \\

EF (\%)
& 59.8 (11) & 60.3 (11)
& 26.0 (21) & 23.0 (14)
& 70.3 (6.8) & 70.2 (7) \\

EDM (g)
& 96.2 (7) & 94.1 (18)
& 103.7 (27) & 103.3 (28)
& 139.3 (18.3) & 139.7 (34.3) \\

ESM (g)
& 93.3 (16) & 95.9 (18)
& 121.3 (14) & 121.7 (15)
& 138.7 (25.5) & 138.7 (27.8) \\


\midrule

\multicolumn{7}{c}{{\textbf{Phase-Contrast Flow2D}}} \\

\cmidrule(lr){2-7}

\textbf{Metric}
& \multicolumn{2}{c}{$\mathbb{C}+\text{Mag}$ ($\mathrm{R}=6$)}
& \multicolumn{2}{c}{$\mathbb{C}+\text{Mag}$ ($\mathrm{R}=8$)}
& \multicolumn{2}{c}{Reference ($\mathrm{R}=1$)}
\\



\midrule

NetV (mL)
& \multicolumn{2}{c}{28.7 (13.4)}
& \multicolumn{2}{c}{28.3 (13.2)}
& \multicolumn{2}{c}{28.7 (12.9)} \\

FV (mL)
& \multicolumn{2}{c}{30.0 (13.1)}
& \multicolumn{2}{c}{29.8 (13.2)}
& \multicolumn{2}{c}{29.2 (12.7)} \\

BV (mL)
& \multicolumn{2}{c}{1.3 (0.85)}
& \multicolumn{2}{c}{1.5 (0.4)}
& \multicolumn{2}{c}{0.5 (0.5)} \\

RF (\%)
& \multicolumn{2}{c}{5.79 (5.8)}
& \multicolumn{2}{c}{6.14 (4.0)}
& \multicolumn{2}{c}{2.4 (2.4)} \\

CO (L/min)
& \multicolumn{2}{c}{1.85 (0.87)}
& \multicolumn{2}{c}{1.83 (0.86)}
& \multicolumn{2}{c}{1.86 (0.84)} \\

\bottomrule
\end{tabular}
\end{adjustbox}
\end{table}

%% file: Tables/ablations.tex
\begin{table}[!b]
\centering
\caption{Ablation study on the reconstruction architecture, DF solver, smoothing operator, and unrolling strategy. The best-performing configuration in each panel is \colorbox{gray!15}{highlighted}, overall best result is shown in \colorbox{green!10}{bold}. ($\spadesuit$) and ($\clubsuit$) denote shared and unshared DF parameters ({\unboldmath ${\lambda}, \mu, \beta$}), respectively.} 
\label{tab:ablation}

\renewcommand{\arraystretch}{1.08}
\setlength{\tabcolsep}{3pt}
\scriptsize

\resizebox{\columnwidth}{!}{%
\begin{tabular}{c c c c c c}
\toprule

$\text{Prox}_{\mathcal{R}(\cdot)}$
& \makecell{DF Solver}
& \makecell{Smoothing}
& Unrolling
& PSNR
& SSIM \\

\midrule

 TE-UNet ($\spadesuit$) & GD       & Quad-S    & ADMM  & 35.60 & 0.936 \\
\cellcolor{gray!15}{TE-UNet ($\clubsuit$)}  & \cellcolor{gray!15}{GD}        & \cellcolor{gray!15}{Quad-S}     &  \cellcolor{gray!15}ADMM &\cellcolor{gray!15}{36.13} & \cellcolor{gray!15}{0.947} \\
ResNet ($\spadesuit$)  & GD         & Quad-S     & ADMM & 35.82 & 0.933 \\
ResNet ($\clubsuit$)  & GD        & Quad-S     & ADMM & 35.83 & 0.935 \\

\midrule

\textit{TE-UNet ($\clubsuit$)}  & GD     & Quad-S     & ADMM &  36.13 & 0.947 \\
\textit{TE-UNet ($\clubsuit$)} & Adam      & Quad-S     & ADMM & 32.79 & 0.868 \\
\textit{TE-UNet ($\clubsuit$)} & Polyak    & Quad-S     & ADMM & 37.30 & 0.954 \\
\cellcolor{gray!15}{\textit{TE-UNet ($\clubsuit$)}} & \cellcolor{gray!15}{Nesterov}  & \cellcolor{gray!15}{Quad-S}     & \cellcolor{gray!15}ADMM & \cellcolor{gray!15}{{37.55}} & \cellcolor{gray!15}{{0.955}} \\

\midrule

\cellcolor{gray!15}{\textit{TE-UNet ($\clubsuit$)}} & \cellcolor{gray!15}{\textit{Nesterov}}  & \cellcolor{gray!15}{Quad-S}     & \cellcolor{gray!15}ADMM & \cellcolor{gray!15}{{37.55}} & \cellcolor{gray!15}{{0.955}} \\
\textit{TE-UNet ($\clubsuit$)}  & \textit{Nesterov} & $\ell_1$-S & ADMM & 35.60 & 0.939 \\
\textit{TE-UNet ($\clubsuit$)} & \textit{Nesterov} & PH-S      & ADMM & 32.92  & 0.913 \\
\textit{TE-UNet ($\clubsuit$)} & \textit{Nesterov} & Log-S      & ADMM & 34.18 & 0.914 \\

\midrule

\cellcolor{green!10}\textbf{\textit{TE-UNet ($\clubsuit$)}} & \cellcolor{green!10}\textbf{\textit{Nesterov}}  & \cellcolor{green!10}\textbf{\textit{Quad-S}} & \cellcolor{green!10}\textbf{ADMM} & \cellcolor{green!10}\textbf{\textbf{37.55}} & \cellcolor{green!10}\textbf{\textbf{0.955}} \\
\textit{TE-UNet ($\clubsuit$)} & \textit{Nesterov}  & \textit{Quad-S} & VSQP & 35.93 & 0.937 \\
\textit{TE-UNet ($\clubsuit$)}  &  \textit{Nesterov} & \textit{Quad-S} & PGD (1 Iter.) & 26.40 & 0.755 \\
\textit{TE-UNet ($\clubsuit$)}  &  \textit{Nesterov} & \textit{Quad-S} & PGD (10 Iters.) & 33.20 & 0.881 \\

\bottomrule
\end{tabular}%
}
\end{table}

%% file: Tables/R1/ablation_R1.tex
\begin{table}[!b]
\centering
\caption{Ablation study on the auxiliary magnitude information. 
The spatial-frequency analysis evaluates the contribution of magnitude information from different relative-error regions using 2D and PE-direction partitions. 
The temporal analysis evaluates the number, distribution, and temporal distance of the auxiliary cardiac phases. Some configurations are unavailable at $\mathrm{R}=6$ because phases separated by $\mathrm{R}$ share the same PE sampling pattern. The \colorbox{green!10}{entries} correspond to $\mathbb{C}\text{+Mag}$ using all available adjacent phases and are included as the reference configuration for comparison, while the best performance in each setting is highlighted in \textbf{bold}.}
\label{tab:ablation_freq_temporal}

\renewcommand{\arraystretch}{1.08}
\setlength{\tabcolsep}{4pt}
\scriptsize


\begin{adjustbox}{max width=\columnwidth}
\begin{tabular}{lcccc}
\toprule
\multicolumn{5}{c}{\textbf{a) Spatial-Frequency Ablation}} \\
\midrule

\multirow{2}{*}{\textbf{Magnitude Region}}
& \multicolumn{2}{c}{\textbf{2D Spatial-Frequency}}
& \multicolumn{2}{c}{\textbf{PE-Direction Spatial-Frequency}} \\

\cmidrule(lr){2-3}
\cmidrule(lr){4-5}

& $\mathrm{R}=6$
& $\mathrm{R}=8$
& $\mathrm{R}=6$
& $\mathrm{R}=8$ \\

\midrule

All ($\Delta$)
& \textbf{40.69 / 0.974}
& \textbf{38.67 / 0.963}
& \textbf{40.69 / 0.974}
& \textbf{38.67 / 0.963} \\

Low Error (1.)
& 39.04 / 0.963
& 37.26 / 0.950
& 39.64 / 0.967
& 37.84 / 0.957 \\

Mid Error (2.)
& 34.88 / 0.930
& 30.75 / 0.868
& 34.80 / 0.931
& 30.14 / 0.855 \\

High Error (3.)
& 34.73 / 0.928
& 30.26 / 0.858
& 34.42 / 0.925
& 30.28 / 0.855 \\

\bottomrule
\end{tabular}%
\end{adjustbox}

\vspace{1pt}

\setlength{\tabcolsep}{9pt}
\begin{adjustbox}{max width=\columnwidth}
\begin{tabular}{lcccc}

\toprule
\multicolumn{5}{c}{\textbf{b) Temporal Ablation}} \\
\midrule

\multirow{2}{*}{\textbf{Setting}}
& \multicolumn{2}{c}{\textbf{$\mathrm{R}=6$}}
& \multicolumn{2}{c}{\textbf{$\mathrm{R}=8$}} \\

\cmidrule(lr){2-3}
\cmidrule(lr){4-5}

& {PSNR}
& {SSIM}
& {PSNR}
& {SSIM} \\

\midrule

\multicolumn{5}{l}{{\colorbox{orange!10}{i) Quantity \& Distribution}}} \\[1pt]

PD-DL ($\mathrm{k}=0$, CG)
& 34.67 & 0.927
& 29.93 & 0.849 \\

$\mathbb{C}+\text{Mag}$ ($\mathrm{k}=2$, Near)
& 37.21 & 0.951
& 32.10 & 0.890 \\

$\mathbb{C}+\text{Mag}$ ($\mathrm{k}=2$, Intermediate)
& 39.00 & 0.963
& 34.53 & 0.921 \\

$\mathbb{C}+\text{Mag}$ ($\mathrm{k}=2$, Far)
& -- & --
& 34.77 & 0.924 \\

$\mathbb{C}+\text{Mag}$ ($\mathrm{k}=4$, Near)
& \textbf{39.63} & \textbf{0.967}
& 35.40 & 0.932 \\

$\mathbb{C}+\text{Mag}$ ($\mathrm{k}=4$, Intermediate)
& -- & --
& 37.14 & 0.948 \\

$\mathbb{C}+\text{Mag}$ ($\mathrm{k}=4$, Far)
& -- & --
& 37.68 & 0.955 \\

$\mathbb{C}+\text{Mag}$ ($\mathrm{k}=6$, Near)
& -- & --
& \textbf{37.93} & \textbf{0.958} \\

\rowcolor{green!10}
\textbf{All adjacent, $\mathrm{k}=\mathrm{R}-1$}
& 40.69
& 0.974
& 38.67
& 0.963 \\

\addlinespace[2pt]
\midrule
\addlinespace[2pt]

\multicolumn{5}{l}{{\colorbox{blue!10}{ii) Temporal Distance}}} \\[1pt]

PD-DL ($\mathrm{k}=0$, CG)
& 34.67 & 0.927
& 29.93 & 0.849 \\

$\mathbb{C}+\text{Mag}$ ($\mathbf{r}_\Delta \in \mathrm{t}+1$)
& \textbf{37.02} & \textbf{0.948}
& \textbf{32.16} & \textbf{0.880} \\

$\mathbb{C}+\text{Mag}$ ($\mathbf{r}_\Delta \in \mathrm{t}+3$)
& 36.25 & 0.944
& 31.25 & 0.866 \\

$\mathbb{C}+\text{Mag}$ ($\mathbf{r}_\Delta \in \mathrm{t}+5$)
& 36.02 & 0.942
& 31.32 & 0.866 \\

\rowcolor{green!10}
\textbf{All adjacent, $\mathrm{k}=\mathrm{R}-1$}
& 40.69
& 0.974
& 38.67
& 0.963 \\

\bottomrule

\end{tabular}%
\end{adjustbox}

\end{table}

%% file: Tables/R1/ablation_compute.tex
\begin{table*}[!t]
\centering
\caption{Computational cost comparison of conventional PD-DL and the different optimization formulations. All methods contain approximately $2.026$ million trainable parameters. Note training times/memory are reported per sample.}
\label{tab:compute_cost}

\renewcommand{\arraystretch}{1.08}
\setlength{\tabcolsep}{7pt}

\begin{tabular}{lcccccccc}
\toprule
& \multicolumn{3}{c}{\textbf{DF Computation}} &
\multicolumn{2}{c}{\textbf{Time (ms)}} &
\multicolumn{2}{c}{\textbf{Peak Memory}} \\
\cmidrule(lr){2-4}
\cmidrule(lr){5-6}
\cmidrule(lr){7-8}

\textbf{Method} &
\textbf{DF Steps} &
$\mathbf{E}/\mathbf{E}^{H}$ \textbf{Evals.} &
\textbf{Parameters} &
\textbf{Inference} &
\textbf{Training} &
\textbf{Inference (MB)} &
\textbf{Training (GB)} \\

\midrule

Conventional PD-DL &
100 & 200 & 2.026M &
$163.9\pm1.1$ &
$526.5\pm18.0$ &
189.2 & 5.8 \\

$\mathbb{C+\text{Mag}}$ (ADMM) &
100 & 400 & 2.026M &
$189.1\pm4.8$ &
$569.6\pm46.9$ &
195.6 & 9.4 \\

$\mathbb{C+\text{Mag}}$ (VSQP) &
100 & 400 & 2.026M &
$187.9\pm5.5$ &
$561.6\pm40.9$ &
195.2 & 9.4 \\

$\mathbb{C+\text{Mag}}$ (PGD, 1 Iter.) &
10 & 40 & 2.026M &
$104.7\pm1.2$ &
$296.9\pm19.5$ &
194.8 & 6.1 \\

$\mathbb{C+\text{Mag}}$ (PGD, 10 Iter.) &
100 & 400 & 2.026M &
$174.9\pm38.4$ &
$534.2\pm241.1$ &
204.7 & 9.4 \\

\bottomrule
\end{tabular}
\end{table*}